\documentclass{aa}

\let\linenumbers\relax

\usepackage[varg]{txfonts}
\usepackage{graphicx} 
\usepackage{gensymb}
\usepackage{color}
\usepackage[colorlinks=true,citecolor=blue,linkcolor=blue]{hyperref}
\usepackage{xspace}
\usepackage{orcidlink}
\usepackage[rightcaption]{sidecap}

\newcommand{\bepposax}{\textsl{Beppo}SAX\xspace}
\newcommand{\rxte}{RXTE\xspace}

\newcommand{\suzaku}{\textsl{Suzaku}\xspace}

\newcommand{\chandra}{\textsl{Chandra}\xspace}
\newcommand{\nustar}{{NuSTAR}\xspace}
\newcommand{\xmmnewton}{{XMM-\textsl{Newton}}\xspace}

\newcommand{\integral}{{INTEGRAL}\xspace}
\newcommand{\xrism}{{XRISM}\xspace}

\newcommand{\newathena}{\textsl{NewAthena}\xspace}
\newcommand{\fermi}{\textsl{Fermi}\xspace}
\newcommand{\tenma}{\textsl{Tenma}\xspace}

\newcommand{\fouru}{4U~1538$-$52\xspace}

\begin{document}

\title{\fouru in a Heartbeat}
\subtitle{Broadband X-ray Spectral Properties from \xmmnewton and \nustar}

\author{
C.~M.~Diez\inst{\ref{affil:IRAP},\ref{affil:ESAC}}\orcidlink{0000-0001-6520-4600}\corrauth{camille.m.diez@gmail.com}\and
S.~Dupourqué\inst{\ref{affil:IRAP}}\orcidlink{0000-0003-2715-8986}\email{sdupourque@irap.omp.eu}\and 
A.-L. Doherty\inst{\ref{affil:NOA},\ref{affil:Crete}}\orcidlink{0009-0009-1816-0356}\email{amie.doherty@noa.gr}\and
D.~Dickson\inst{\ref{affil:Leuven}}\orcidlink{0009-0006-4860-9212}\email{davey.dickson@kuleuven.be}\and 
A. Zainab\inst{\ref{affil:remeis}}\orcidlink{0000-0003-1163-9320}\email{aafiazainab.ansar@fau.de}\and
S. Martínez-Núñez\inst{\ref{affil:IFCA}}\orcidlink{0000-0002-5134-4191}\email{smartinez@ifca.unican.es}\and
J.~M. Torrej\'{o}n\inst{\ref{affil:IUFACYT}}\orcidlink{0000-0002-5967-5163}\email{jmt@ua.es}\and
S. Guillot\inst{\ref{affil:IRAP}}\orcidlink{0000-0002-6449-106X}\email{sebastien.guillot@utoulouse.fr}\and
A.~Rouco~Escorial\inst{\ref{affil:Starion}}\orcidlink{0000-0003-3937-0618}\email{a.roucoescorial@stariongroup.eu}\and
F. Fürst\inst{\ref{affil:ESAC}}\orcidlink{0000-0003-0388-0560}\email{felix.fuerst@esa.int}
\and
P. Kretschmar\inst{\ref{affil:ESAC}}\orcidlink{0000-0001-9840-2048}\email{peter.kretschmar@esa.int}
}

\institute{Univ Toulouse, CNES, CNRS, IRAP, Toulouse, France \label{affil:IRAP}
\and European Space Agency (ESA), European Space Astronomy Centre (ESAC), Camino Bajo del Castillo s/n, 28692 Villanueva de la Cañada, Madrid, Spain \label{affil:ESAC}
\and Institute for Astronomy \& Astrophysics, National Observatory of Athens, V. Paulou \& I. Metaxa, 11532 Athens, Greece \label{affil:NOA}
\and Department of Physics, University of Crete, 71003 Heraklion, Greece \label{affil:Crete}
\and Institute of Astronomy, KU Leuven,
Celestijnenlaan 200D, 3001, Leuven, Belgium \label{affil:Leuven}
\and Dr. Karl Remeis Observatory, Bamberg and Erlangen Center for Astroparticle Physics, Friedrich Alexander University Erlangen-Nuernberg, Germany \label{affil:remeis}
\and Instituto de Física de Cantabria (CSIC-UC), Av. de los Castros s/n, 39005 Santander, Spain \label{affil:IFCA} 
\and Instituto Universitario de F\'{\i}sica Aplicada a las Ciencias y las Ingenier\'{\i}as, Universidad de Alicante, 03690 Alicante, Spain \label{affil:IUFACYT}
\and Starion España S.L.U, calle Chile 10, oficina 247, 28290 Las Rozas, Madrid, Spain \label{affil:Starion}
}
\date{Received 7 July 2026 / Accepted 12 August 2026}

\abstract{Galactic high-mass X-ray binaries (HMXBs) are important systems for studying accretion mechanisms onto compact objects and for investigating the complex stellar winds of massive stars. In particular, HMXBs hosting a neutron star allow us to reveal the structure of the accreted material in X-ray pulsars and consequently to investigate how matter behaves under extreme conditions of pressure and density. These are major scientific goals for \xrism and \newathena.}{Here we report on the first out-of-eclipse \xmmnewton observation of the HMXB \fouru, complemented by \nustar coverage. Our campaign aimed to investigate stellar-wind variability and continuum changes with high-resolution spectroscopy at a critical orbital phase: when the neutron star is in inferior conjunction.}{Thanks to simultaneous observations covering both soft and hard X-rays, we obtain the most detailed X-ray view of the accreted material in \fouru to date. In particular, we perform time-resolved spectroscopy down to the pulse period of the neutron star to highlight wind clumping properties and accretion structures.}{In this dataset, we observe a bright flare reaching $\sim$10$^{37}\,\rm{erg\,s^{-1}}$ probably induced by the accretion of a $10^{20}\,\mathrm{g}$ clump, followed by a luminosity dip forming a heartbeat-like episode. This event is followed by three local absorption peaks with local variability of the order of the pulse period, and a gradual hardening of the underlying spectrum throughout the observation. This could indicate the presence of both small-scale and large-scale overdense structures in the vicinity of the neutron star, which can be attributed to clumps and filamentary structures embedded in the accretion wake. These observational evidences are further supported and reproduced by 3D hydrodynamic simulations.}{} 

\keywords{X-rays: binaries,  stars: neutron,  stars: winds, outflows}

\maketitle

\section{Introduction}
\label{section:intro}

\object{4U 1538-52} is a bright \citep[$L_{\rm{0.3-11.5\,keV}} \approx 10^{36}\,\rm{erg\,s^{-1}}$,][]{Rodes-Roca_2011} eclipsing persistent neutron star high-mass X-ray binary (HMXB) located at $\sim$5.6$^{+0.5}_{-0.4}$\,kpc \citep{Bailer_Jones21}. The X-ray pulsations from the slow-rotating pulsar were first discovered by \cite{Davison_1977} and \cite{Becker_1977} using \textsl{Ariel} 5 and OSO-8 data with an estimated pulse period of $\sim$528\,s. The neutron star orbits its B0Iab supergiant companion, QV Nor \citep{Reynolds_1992} in a close, mildly eccentric 3.72\,d orbit \citep{Hemphill_2019} with $a \sin i = 53.1\pm1.5\,\textrm{lt-s}$ and $e \lesssim 0.2$ \citep{Mukherjee_2006}. The neutron star is deeply embedded in the stellar wind, moving at distances between $\sim$1.35 and $\lesssim$1.5 stellar radii \citep{Torrejon_2015}. Given these convenient parameters, we can probe a wide range of orbital phases in a relatively small amount of time. 

\fouru appears similar in its properties to the well-studied HMXB Vela X-1 \citep[$e = 0.0898\pm0.0012$, at a typical distance $\sim$1.8 stellar radii,][]{Kretschmar_2021a}. However, in contrast to the marked changes observed in Vela X-1 \citep{Kretschmar_2021a,Diez_2023}, observations of \fouru with \rxte/PCA \citep{Jahoda_2006} and \bepposax \citep{Boella_1997} revealed a stable absorption column density $N_{\rm{H}}$ along the orbit \citep{Mukherjee_2006}. On the other hand, \suzaku observed \fouru at $\phi_{\rm{orb}} \approx 0.50$--$0.68$ \citep{Hemphill_2014} (orbital phases updated using the ephemeris by \citealp{Hemphill_2019}) where they measured $N_{\rm{H}}$ spikes ($\sim$~$10^{23}\, \rm{cm^{-2}}$) towards the end of the observation. They attribute these changes to the accretion of an overdense region of a clumpy stellar wind.

Accretion onto magnetised neutron stars is usually split into sub- and super-critical accretion regimes separated by the critical luminosity $L_\mathrm{crit}$ as first described in \citet{Basko_Sunyaev_1976}. Below $L_\mathrm{crit}$, the infalling material is not stopped by radiation pressure and is therefore halted very near the surface of the neutron star, creating hotspots, also called thermal mounds  \citep{Zeldovich_Shakura_1969}. In a sub-critical accretion regime, the X-rays are likely to be emitted in an upward direction, forming a ``pencil'' beam aligned with the magnetic field lines \citep{Basko_Sunyaev_1976, Meszaros_1984, Mushtukov_Tsygankov_2022}. Alternatively, above $L_\mathrm{crit}$ in the super-critical accretion regime, the radiation pressure is strong enough to decelerate the infalling material well above the surface and a radiation-dominated shock forms. In such a configuration, the observed X-rays escape sideways from the accretion column in a ``fan'' beam perpendicular to the magnetic field lines \citep[e.g.][]{Farinelli_2016}. 

The spectrum of \fouru exhibits several features typical of accreting neutron stars. A fluorescent emission line at $\sim$6.4~keV, first identified by \citet{Makishima_1987} with \tenma and more recently seen with \chandra and \xmmnewton data \citep[e.g.,][]{Torrejon_2010, Gimenez_2015}, is associated with the Fe~K$\alpha$ complex. Soft emission lines below 3 keV have been identified by \citet{Torrejon_2015} and \citet{Rodes-Roca_2011} using \chandra and \xmmnewton data taken during the eclipse. The presence of these soft features is a strong indicator of an extended ionised inhomogeneous wind in which the neutron star is embedded \citep[e.g.,][]{Rodes-Roca_2011}. Two broad absorption features are often observed in \fouru: one at $\sim$22 keV and another at $\sim$50 keV. They have been identified as cyclotron resonant scattering features (CRSFs, or cyclotron lines), corresponding to the fundamental \citep{Clark_1990} and its first harmonic \citep{Robba_2001}, respectively. Cyclotron lines appear as broad absorption features in the spectra of strongly magnetised neutron stars. They allow a direct measurement of the local magnetic field strength through the energy at which they appear. The fundamental CRSF originates from resonant scattering of photons by electrons that are lifted from the ground state to the first excited Landau level and subsequently decay radiatively \citep[for an overview, see][and references therein]{Staubert_2019}. Transitions involving higher Landau levels produce the harmonic lines. In addition, the evolution of CRSF features with the luminosity of the source can probe accretion geometry onto the magnetised neutron star \citep[see reviews of the field in][]{Mushtukov_Tsygankov_2022, Saathoff_PhD}.

\begin{figure}[htpb!]
    \centering
    \centerline{\includegraphics[trim=0cm 0cm 0cm 0cm, clip=true, width=1.0\linewidth]{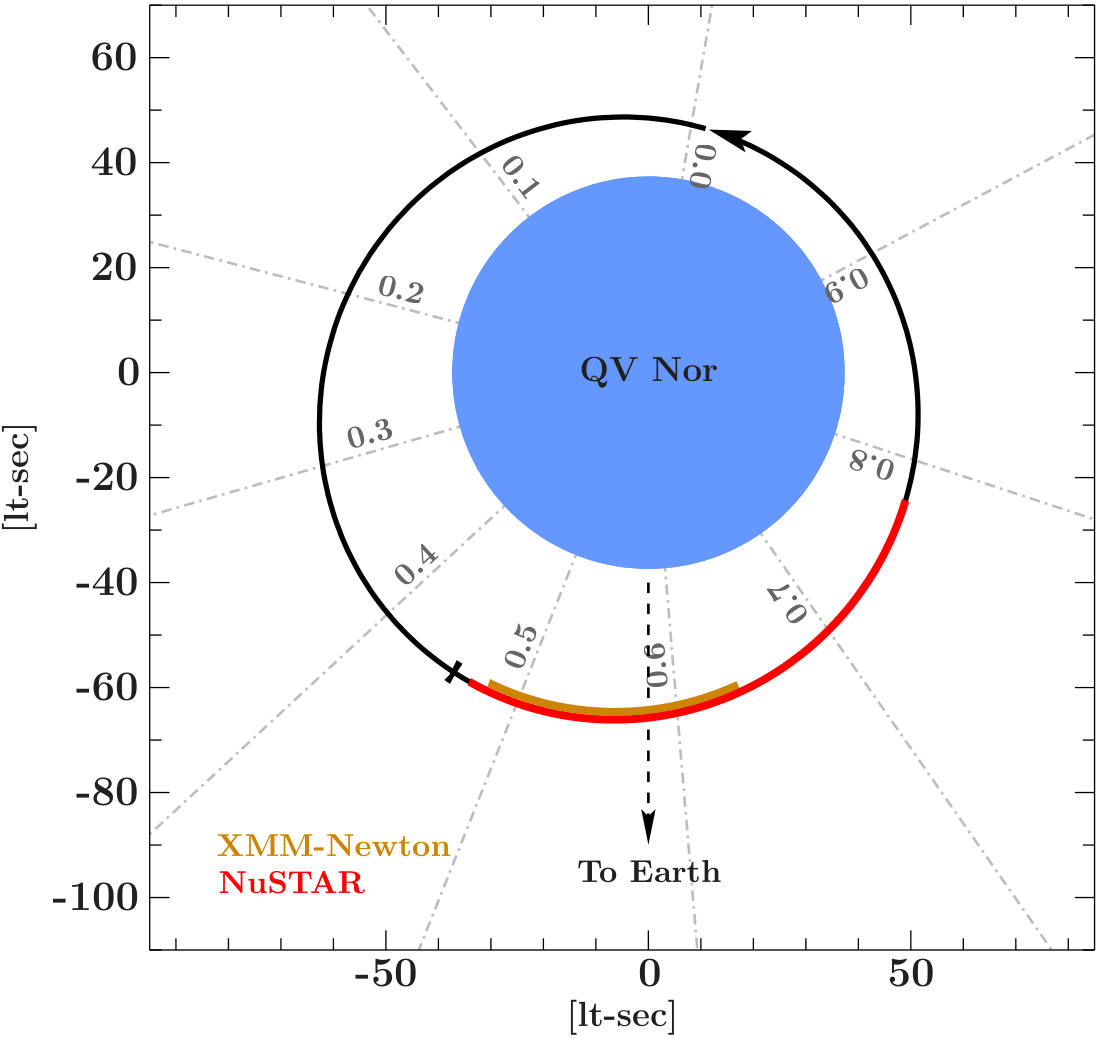}}
    \caption{Sketch of the \fouru system showcasing the orbital phases covered by the neutron star during our \xmmnewton (gold) and \nustar (red) observations. In this image, the observer is facing the system at the bottom of the picture, and $\phi_{\rm{orb}} = 0$ is defined at the mid-eclipse time. The apastron is indicated by the black tick around $\phi_{\rm{orb}} = 0.46$.}
    \label{fig:orbit}
\end{figure}

Our observational campaign aimed at investigating absorption changes with high resolution provides the first out-of-eclipse \xmmnewton observation of \fouru, complemented by \nustar coverage. Together, these two instruments give us a broadband energy coverage of \fouru with high-time resolution, allowing us to study the stellar wind variability and changes in the continuum at different timescales. Aiming to maximise the brightness of the source while probing the dense regions of the wind at the same time, we targeted orbital phases wherein the neutron star is at inferior conjunction (see Fig.~\ref{fig:orbit}). Such an approach was previously successfully applied to another classical HMXB, Vela X-1, and provided insights on the accretion geometry onto the neutron star \citep{Diez_2022} and an unprecedented detailed view of the structure and chemical composition of the stellar wind surrounding the neutron star \citep{Diez_2023, Diez_2025}.

In this paper, we examine our joint \xmmnewton and \nustar observation of \fouru in detail. We first describe the two datasets, the software used, and the specific data-reduction methods in Sect.~\ref{section:obs_data_reduc}. We then present the light curves and timing results in Sect.~\ref{section:timing}. We detail our spectral modelling method in Sect.~\ref{section:spectral_modelling} and subsequent time-resolved spectroscopy in Sect.~\ref{section:spectral_analysis}. We discuss our results in Sect.~\ref{section:discussion} and give a conclusion of our work with an outlook in Sect.~\ref{section:conclusion}.

\section{Observation and data reduction}
\label{section:obs_data_reduc}

\begin{table*}
\renewcommand{\arraystretch}{1.1}
\caption{Observations log.}
\label{tab:obs_details}
\begin{center}
\begin{small}
\begin{tabular}{llcccccc}    
\hline\hline
Mission &
Instrument &
Obs ID & 
\multicolumn{1}{c}{Time Start} & 
\multicolumn{1}{c}{Time Stop} &
\multicolumn{1}{c}{Time duration} &
\multicolumn{1}{c}{On-source exposure} &
\multicolumn{1}{c}{Orbital phase} \\
&
&
&
\multicolumn{1}{c}{MJD (day)$^{(a)}$} & 
\multicolumn{1}{c}{MJD (day)$^{(a)}$} &
\multicolumn{1}{c}{(ks)} &
\multicolumn{1}{c}{(ks)} &
\multicolumn{1}{c}{(with $T_{\mathrm{ecl}}$)} \\
\hline
\xmmnewton & EPIC-pn & {0951810101}     & 60555.8182     & 60556.4214     & 52.125 & 36.534$^{(b)}$      & 0.478--0.639          \\
\nustar & FPMA/B & {31001016002}     & 60555.7786     & 60556.9448  & 100.734 & 55.442      & 0.467--0.780          \\
\hline
\multicolumn{8}{p{0.8\linewidth}}{$^{(a)}$ The start and stop times are given as modified Julian dates (MJD).} \\
\multicolumn{8}{p{0.8\linewidth}}{$^{(b)}$ The live time fraction is of 71\% when using EPIC-pn in small window mode.}
\end{tabular}
\end{small}
\end{center}
\renewcommand{\arraystretch}{1.0}
\end{table*}

\fouru was observed on 2 September 2024 as the main science target of a simultaneous \nustar, and \xmmnewton campaign during the \nustar GO Cycle 10. On board \xmmnewton \citep{Jansen_2001}, we use the European photon imaging camera pn-CCDs \citep[EPIC-pn;][]{Strueder_2001}, the EPIC metal oxide semi conductor \citep[EPIC-MOS;][]{Turner_2001}, and the reflection grating spectrometers \citep[RGS;][]{denHerder_2001} under OBSID 0951810101. On board \nustar \citep{Harrison_2013}, we use the focal plane modules A and B (FPMA and FPMB) under OBSID 31001016002. For this work, we only exploit \xmmnewton/EPIC-pn and \nustar/FPMA-B data. Details of the observation are given in Table~\ref{tab:obs_details}. As expected due to the low Earth orbit (LEO) of \nustar, the \xmmnewton net exposure time is larger than the \nustar one. The orbital phases $\phi_{\rm{orb}}$ mentioned in this work are derived from the ephemeris in \citet{Hemphill_2019}, and we define $\phi_{\rm{orb}} = 0$ at the mid-eclipse time.

All the times mentioned in this work are barycentred and corrected from the influence of the double-star motion for circular or elliptical orbits (i.e. binary-corrected). The spectra are rebinned using the optimal rebinning approach from \citet{Kaastra_2016} with the \texttt{ftgrouppha} ftool from the HEASOFTv6.35.1 package. For the spectral analysis, we use \texttt{jaxspec v0.3.4} \citep{Dupourque_2024}, a Python library for X-ray spectral analysis with Bayesian inference via Markov Chain Monte Carlo (MCMC) sampling using the No U-Turn Sampler \citep[NUTS, ][]{nuts_sampler}. We use the NUTS implementation from \texttt{numpyro} \citep{numpyro}, running 20 chains with $10^3$ warmup and sampling steps running on a single Nvidia H100 GPU, and convergence was asserted using the $\hat{R}<1.01$ criterion \citep{rhat}. The reliability and efficient convergence of the NUTS sampler allows us to thaw all the parameters of the fit in Sect.~\ref{section:strict_simult_xmm_nustar} (except the ones used as reference values) and to obtain robust constraints on the parameters within short timescales for high-time-resolved spectroscopy in Sect.~\ref{section:gti_spectro} and \ref{section:pulse_spectro}. This method was recently successfully applied to Vela X-1 for simultaneous \xrism, \nustar and \xmmnewton data \citep{Diez_2025}. For comparison and sanity checks with \texttt{jaxspec}, we also use the Interactive Spectral Interpretation System (\texttt{ISIS}) v1.6.2-53 \citep{Houck&Denicola_2000}. Both \texttt{ISIS} and \texttt{jaxspec} provide access to \texttt{XSPEC} \citep{Arnaud_1996} models, which are referenced later in the text. Errors are given within the 2-$\sigma$ confidence level, unless otherwise stated.  

\subsection{\nustar}
\label{section:nustar}

We use the NuSTARDAS pipeline (\texttt{nupipeline}) v2.1.5 with associated CALDB v20241015 applied with the clock correction to extract spectra and light curves. We select the source region with a radius of $80''$ centred on the PSF peak. The background region is selected with a radius of $60''$ as far away from the source as possible, since \fouru is so bright that it illuminates most of both focal planes. This method, however, introduces systematic uncertainties due to variations in the background across the instrument's field of view. We decide to ignore energies higher than 50\,keV due to the low signal-to-noise ratio (S/N) permitted by \nustar at higher energies, combined with the moderate luminosity of the source. Thus, the source remains approximately two times brighter than the background even at the highest energies considered during this study, thus minimising the impact of remaining uncertainties. An unexplained excess/upturn in the \nustar data at low energies \citep[see, e.g.,][]{Akylas_2019, Diez_2023} is also observed here in the energy range where \nustar overlaps with \xmmnewton. We thus decide to ignore \nustar energies below 5 keV for our spectral analysis to mitigate this effect.

\subsection{\xmmnewton}
\label{section:xmm}

For this work, we analyse data from EPIC-pn only. The EPIC-MOS spectra are too severely affected by pile-up \citep{Jethwa_2015}, a phenomenon caused by multiple photons hitting adjacent pixels during a single read-out interval. As a result, these photons are erroneously recorded as single events with an energy corresponding to the combined energies of the individual photons. This effect strongly modifies the spectral shape. Hence, we do not consider the EPIC-MOS data for our analysis, as is often the case for bright sources \citep[e.g., during the outburst of the ms X-ray pulsar SAX J1808.4-3658 in][]{Papitto_2009}.

\begin{figure}[htbp!]
    \centering
    \includegraphics[trim=2cm 5cm 1cm 7cm, clip=true, width=1.0\linewidth]{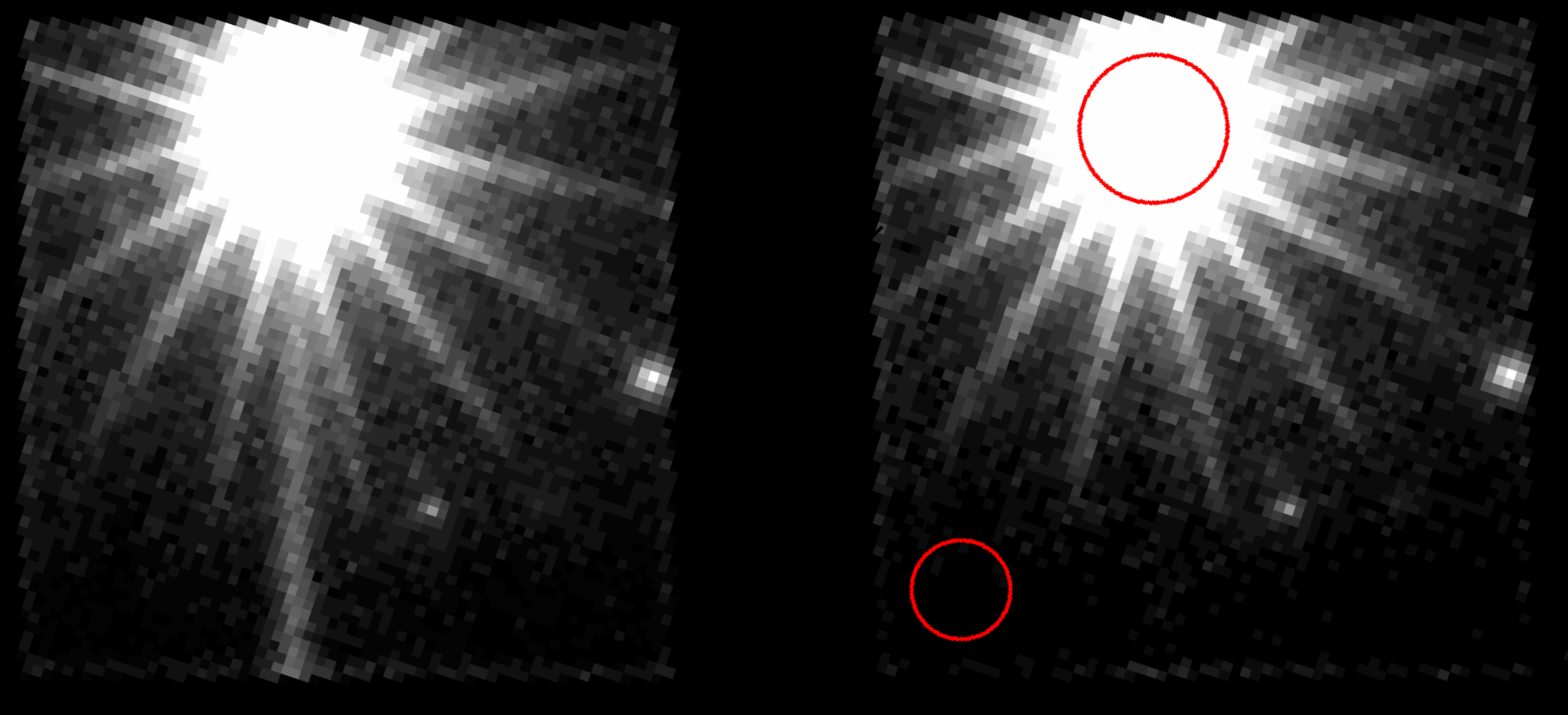}
    \caption{Before (left figure) and after (right figure) out-of-time events correction for the \xmmnewton/EPIC-pn data. The source and background regions are highlighted by the red circles.}
    \label{fig:oot_corr}
\end{figure}

The observation with EPIC-pn was set up in small window mode with a thin filter. We use the Science Analysis System software v22.1.0 with the Current Calibration Files as of July 2025. We start from the Observation Data Files level and run \texttt{epproc}. We select the source region with a radius of $30''$ centred on the PSF peak and a background region with a radius of $20''$ as far from the source as possible (see Fig.~\ref{fig:oot_corr}). Correction for the flaring particle background is unnecessary given the source's high brightness, and events on bad pixels are excluded. We correct for out-of-time (OoT) events that are registered during the readout of a CCD. These events can broaden the spectral features and create a strip of events with wrongly reconstructed positions that is visible in the image file (Fig.~\ref{fig:oot_corr}). To correct for this, we extract an OoT event list and a spectrum file from it. We then multiply the values in the column \texttt{CTS\_OOT} by the expected fraction of OoT events (1.1\% for EPIC-pn in small window mode\footnote{\url{https://xmm-tools.cosmos.esa.int/external/xmm\_user\_support/documentation/uhb/epicoot.html}}). We then subtract the rescaled values of the \texttt{CTS\_OOT} from the \texttt{COUNTS} column of the source spectrum. We finally analyse events in the 0.5-10 keV band, where the S/N is optimal.

\section{Light curve and timing}
\label{section:timing}

\begin{SCfigure*}[1.0]
    \includegraphics[trim=0cm 0cm 0cm 0.75cm, clip=true, width=1.45\linewidth]{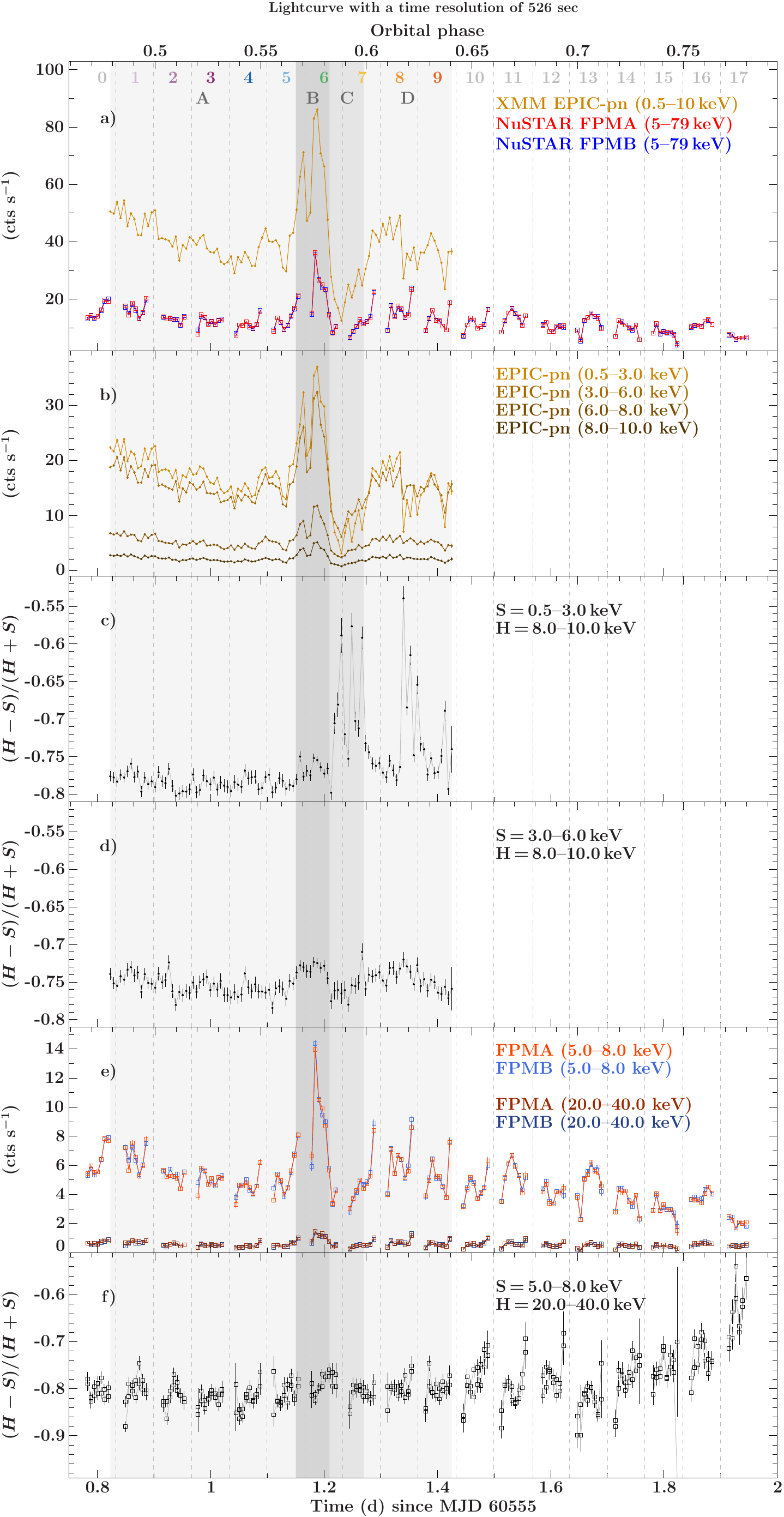}
    \centering{\caption{\textit{Panel a)}: Light curves (count rates) for \xmmnewton/EPIC-pn (gold), \nustar/FPMA (red) and FPMB (blue). \textit{Panel b)}: Energy-resolved \xmmnewton/EPIC-pn light curves. \textit{Panels c)} and \textit{d)}:
    \xmmnewton/EPIC-pn hardness ratios as $(H-S)/(H+S)$ in the mentioned energy bands. \textit{Panel e)}: Energy-resolved \nustar light curves. \textit{Panel f)}: \nustar hardness ratios as $(H-S)/(H+S)$ in the mentioned energy bands. For all panels, squares are for \nustar data, points are for \xmmnewton data. The time resolution is of $P_{\rm{spin}}=526.03\, \mathrm{s}$ and the data are plotted along the time (bottom axis) and the orbital phase (top axis). The shaded regions named A, B, C and D highlight the pre-flare, flare, dip and post-dip periods, when coverage from both \xmmnewton and \nustar is available. The dashed vertical lines separate the numbered GTIs we chose based on the length of a \nustar exposure cycle for which we perform the corresponding time-resolved spectroscopy in Sect.~\ref{section:gti_spectro}.}}
    \label{fig:hr}
\end{SCfigure*}

We show in \textit{panel a} of Fig.~\ref{fig:hr} the \xmmnewton/EPIC-pn and \nustar light curves binned at the pulse period of the neutron star to filter out the variability caused by the source pulsations. To compute the pulse period $P_{\rm{spin}}$ of \fouru, we use the $Z^2$ statistics \citep{Buccheri_1983_Z2} based on photon events collected from all instruments used in this work. We correct the events for both barycentric motion and binary orbital effects with the ephemeris from \citet{Hemphill_2019}. The results are consistent, yielding a period of $P_{\rm{spin}} = 526.03\pm0.04\,\mathrm{s}$, in agreement with the source pulse period history from the \fermi Gamma-ray Burst Monitor{\footnote{\url{https://gammaray.nsstc.nasa.gov/gbm/science/pulsars/lightcurves/4u1538.html}}}.

In \textit{panel a} of Fig.~\ref{fig:hr}, we observe a major flare with both observatories at $T_{\mathrm{obs}} \approx 60556.19$ MJD reaching $\sim$86~$\mathrm{counts \ s^{-1}}$ with EPIC-pn, which is more than two times higher than the average count rate. This flare is quickly followed 0.03 d ($\sim$43 min) later by a major dip going down to $\sim$12~$\mathrm{counts \ s^{-1}}$ with EPIC-pn. This post-flare dip is more than three times lower than the average count rate, producing a distinct heartbeat-shaped episode. We highlight in Fig.~\ref{fig:hr} four segments corresponding to the pre-flare, flare, dip and post-dip portions when coverage from both \xmmnewton and \nustar is available, labelled A, B, C and D respectively. The regular gaps observed in the \nustar light curve correspond to the Earth's passage in the \nustar field of view. We indicate each \nustar exposure cycle covered during this observation by numbers in Fig.~\ref{fig:hr}, and we perform the corresponding time-resolved spectroscopy in Sect.~\ref{section:gti_spectro} to investigate further spectral variability.

The presence of a clumpy absorber shows signatures in the soft X-rays as highlighted, e.g., in the \suzaku and \xmmnewton light curves analysed by \citet{Hemphill_2014} and \citet{Rodes-Roca_2011}. In an effort to quantitatively study the local absorber and the intrinsic variability of \fouru at low energies, we present in Fig.~\ref{fig:hr} the energy-resolved light curves and corresponding hardness ratios (HR) derived from the \xmmnewton/EPIC-pn observation. The HR is defined as $\mathrm{HR} = (\rm{Hard} - \rm{Soft})/(\rm{Hard}+\rm{Soft})$, between count rates in the energy bands labelled in the figure. In \textit{panels b} and \textit{e} of Fig.~\ref{fig:hr}, we notice the presence of the flare, marked by the shaded region around $\phi_{\mathrm{orb}}\approx 0.57$, in all energy bands but the strong dip following the flare, marked by the shaded region around $\phi_{\mathrm{orb}}\approx 0.59$, is visible mostly in the softest bands: $0.5$--$3$ keV and $3$--$6$ keV for \xmmnewton, and $5$--$8$ keV for \nustar. In \textit{panel c} of the same figure, we observe a clear hardening of the underlying spectrum during the dip, as well as later at $\phi_{\mathrm{orb}} \approx 0.62$, corresponding to a local (albeit less pronounced) dip in the EPIC-pn light curve. A weak hardening is also visible in \textit{panel d}, coinciding with the flare. However, variations in HR are the most pronounced when the softest energy band (0.5–3 keV) is included. This pattern is characteristic of changes driven by absorbing material, which primarily affects low energies, whereas continuum variations from the neutron star impact all energy bands. When including count rates from a harder band as possible with \nustar in \textit{panel f}, we observe a gradual increase of the HR towards the end of the observation. 
 

\section{Spectral modelling}
\label{section:spectral_modelling}

The HR variability displayed by \fouru requires a spectral analysis of the data at different timescales. To determine the best-fit model to describe these datasets, we start by selecting \nustar and \xmmnewton data that are strictly simultaneous, i.e., between 60555.82 MJD and 60556.43 MJD. This allows us to define the best-fit model using the full available energy range and accounting for variability observed by both instruments simultaneously. Known cross-calibration issues have been reported between \xmmnewton and \nustar \citep[e.g.,][]{Gokus_2017, Diez_2023}, but also between both FPMs onboard \nustar \citep[e.g.,][]{Zalot_2024}. We thus define three floating flux cross-normalisation constants $\mathrm{\mathcal{C}_{FPMA}}$, $\mathrm{\mathcal{C}_{FPMB}}$ and $\mathrm{\mathcal{C}_{EPIC-pn}}$ to give the relative normalisation between all instruments, fixing $\mathrm{\mathcal{C}_{FPMA}}$ to 1 as a reference. To allow for energy calibration shifts and uncertainties between the different detectors, we introduce three offset constants $\mathcal{S}_{\rm{FPMA}}$, $\mathcal{S}_{\rm{FPMB}}$ and $\mathcal{S}_{\rm{EPIC-pn}}$, fixing $\mathcal{S}_{\rm{FPMA}}$ to 0\,keV as the reference. We address model variability and track changes in spectral parameters when performing time-resolved spectroscopy in Sect.~\ref{section:spectral_analysis}. 

\subsection{Continuum and partial coverer}

To facilitate direct comparisons and ensure homogeneous continuity of the results, we employ a phenomenological approach that is commonly done for \fouru \citep{Rodes-Roca_2009,Hemphill_2014, Hemphill_2016, Hemphill_2019}, using a power law modified by a high-energy cutoff. Two main flavours of high-energy cutoffs are commonly found in the literature to describe the continuum of bright HMXBs (see, e.g., \citealp{Hemphill_2014, Hemphill_2019} for \fouru or \citealp{Fuerst_2014a, Diez_2022} for Vela X-1): \texttt{Highecut} \citep{White_1983} and \texttt{FDcut} \citep{Tanaka_1986}. Following the nomenclature of \texttt{jaxspec}, they can be described with the cutoff $E_{\rm{cut}}$ and the folding $E_{\rm{fold}}$ energies, both in keV, as:

\begin{equation*}
\texttt{Powerlaw}\times\texttt{Highecut} = K\, E^{-\Gamma} \times 
\begin{cases} 
\exp \left( \frac{E_{\rm{cut}} - E}{E_{\rm{fold}}} \right)& \text{if $E > E_{\rm{cut}}$}\\ 1 & \text{if $E \leq E_{\rm{cut}}$}
\end{cases}
\end{equation*}
\begin{equation*}
\texttt{Powerlaw}\times\texttt{FDcut}(E) = K\,E^{-\Gamma} \left[ 1 + \exp\left( \frac{E - E_{\text{cut}}}{E_{\text{fold}}} \right) \right]^{-1},
\end{equation*}
where $K$ and $\Gamma$ stand for the normalisation $[\rm{ph \ keV^{-1} \ cm^{-2} \ s^{-1}}]$ at 1 keV and the photon index of the power law, respectively.

The fit yields a slightly worse description of the data with \texttt{FDcut} than with \texttt{Highecut}. We also test a simple cutoff power law \texttt{Cutoffpl} as in \citet{Hemphill_2013}, which yields residuals at high energies and is discarded. Furthermore, we test \texttt{npex} \citep{Mihara_1995} as in \citet{Hemphill_2014, Hemphill_2019}, but the second power law introduces additional degrees of freedom. This complicates the interpretation of the fit and results in broad, deep cyclotron lines that reach the hard limits of the priors. \citet{Rodes-Roca_2011} used a three-power-law continuum, but the physical interpretation remains limited.

To account for residuals at low energies caused by the obscuration of the source by the interstellar medium (ISM) with equivalent hydrogen column density $N_{\rm{H,ISM}}$, and varying local absorption $N_{\rm{H}}$ in the vicinity of the neutron star, we further modify the continuum with a partial-covering absorption component \citep[e.g.,][]{Hemphill_2014, Torrejon_2015}: 
\begin{equation*}
\texttt{Tbabs}_{\rm{ISM}} \, \times \, \texttt{Tbpcf} = \exp^{-N_{\rm{H,ISM}}\,\sigma(E)}\,[{\rm{CF}} \, \times \, \exp^{-N_{\rm{H}}\,\sigma(E)}+\,(1-{\rm{CF}})]
\end{equation*}

We leave $N_{\rm{H}}$ free to vary and $N_{\rm{H,ISM}}$ is set to $7.33 \times 10^{21}\,\rm{cm^{-2}}$ following the value obtained from the NASA HEASARC $N_{\rm{H}}$ tool website \citep{HI4PI_2016}. Abundances are from \cite{Wilms_2000} and cross-sections $\sigma(E)$ from \cite{Verner_1996}. The covering fraction $\rm{CF}$ provides a measure of the inhomogeneity of the obscurer by indicating how much of the X-ray light is covered. It varies from $0$ (no source coverage) to $1$ (full source coverage). The partial covering model stands out as the most effective approach for characterising the absorber in \fouru, as demonstrated in three different continuum models \citep{Hemphill_2014}, and also in other obscured accreting X-ray pulsars such as 4U 0115+63 \citep{Farinelli_2016} or Vela X-1 \citep[e.g.,][]{Diez_2023}. 

\subsection{Absorption and emission lines}
Thanks to the high-energy coverage permitted by \nustar, we confirm once again the presence of a fundamental cyclotron line at $\sim$21\,keV (see hereafter Sect.~\ref{section:strict_simult_xmm_nustar} and Table~\ref{tab:best_fit_strict_simult}) in the spectrum of \fouru. We can observe residuals around $\sim$46\,keV, possibly corresponding to the first harmonic of the cyclotron line, but its clear detection in this dataset is limited by the low S/N at these energies; thus, we can only obtain a lower limit on its line energy. We attribute the residuals in absorption around $\sim$9\,keV (see hereafter  Sect.~\ref{section:strict_simult_xmm_nustar}, Table~\ref{tab:best_fit_strict_simult}, and Fig.~\ref{fig:spec_strict_simult}) to the presence of a possible 10-keV feature in the spectrum of \fouru. \citet{Coburn_2001} first reported this feature in emission. Later, \citet{Hemphill_2013} also observed this feature in emission around $\sim$11\,keV using a \texttt{cutoffpl} continuum model in \integral observations spanning 7 years. The \rxte spectra analysed in \citet{Hemphill_2016} required an absorption component around $\sim$8 keV (or possibly emission around $\sim$12 keV). On the other hand, \citet{Manikantan_2023} do not observe this feature with a \texttt{npex} continuum model on an 11-ks \nustar observation. The discrepancies between our work and previous studies may lie in the different continuum models used, a reduced exposure time in the case of \citet{Manikantan_2023}, non-continuous observations spread over a long time period, and the result of data from different instruments. The origin of this feature is still unknown and is likely due to the simplistic nature of current continuum models for understanding the complex underlying accretion physics \citep{Coburn_2002}. A deeper investigation of the 10-keV feature is out of the scope of this paper. We model the absorption features present in our dataset with three multiplicative broad Gaussian absorption components \texttt{Gabs} such as:
\begin{align*}
\texttt{Gabs} = \exp\left[\frac{-2 \times d_{\rm{gabs}}}{\sigma_{\rm{gabs}}\sqrt{2\pi}\left[1-{\rm{erf}}\left(\frac{-E_{\rm{gabs}}}{\sigma_{\rm{gabs}}\sqrt{2\pi}}\right)\right]}\exp\left(\frac{-(E-E_{\rm{gabs}})^2}{2\sigma_{\rm{gabs}}^2}\right)\right],
\end{align*}
with $d_{\rm{gabs}}$ and $\sigma_{\rm{gabs}}$ the feature depth and width both in keV. 

Thanks to the good spectral resolution enabled by \xmmnewton, we detect three emission components at $\sim$0.98\,keV, $\sim$1.64\,keV, and $\sim$6.39\,keV that we model with additive Gaussian line profiles so that:
\begin{align*}
\texttt{Gauss} = \frac{2\times A_{\rm{L}}}{\sigma_{\rm{L}}\sqrt{2\pi}\left[1-{\rm{erf}}\left(\frac{-E_{\rm{L}}}{\sigma_{\rm{L}}\sqrt{2}}\right)\right]} \exp\left(\frac{-(E-E_{\rm{L}})^2}{2\sigma_{\rm{L}}^2}\right),
\end{align*}
where $A_{\rm{L}}$ is the total area [$\rm{ph \ cm^{-2} \ s^{-1}}$] in the line,  $E_{\rm{L}}$ the line energy and $\sigma_{\rm{L}}$ the line width both in keV. 

The emission feature at $\sim$6.39\,keV is naturally associated with the fluorescent near-neutral Fe K$\alpha$ complex as commonly observed in this source \citep[e.g.,][]{Rodes-Roca_2011, Torrejon_2015, Hemphill_2014, Hemphill_2019}. Residuals below 3\,keV are likely due to unresolved and blended emission lines originating from the ionised plasma surrounding the X-ray source, a characteristic feature of wind-accreting systems \citep[e.g.,][]{Torrejon_2015, Diez_2023}. We attribute the Gaussian component at $\sim$0.98\,keV to the blended H-like Ne~X Ly$\alpha$ and He-like Ne~IX complex \citep{GarciaMack_1965, Yerokhin_2019}. Similarly, the residuals around $\sim$1.64\,keV are consistent with the ionised H-like Mg~XII Ly$\alpha$ and He-like Mg~XI complex \citep{GarciaMack_1965, Yerokhin_2019}. However, contributions from neighbouring near-neutral Si~K$\alpha$ and ionised Si~XIII lines \citep{Hell_2016, Yerokhin_2019} cannot be excluded due to the limited energy resolution of EPIC-pn. These features are consistent with the Mg and Si emission lines previously reported in \fouru using \chandra \citep{Torrejon_2015} and \xmmnewton \citep{Rodes-Roca_2011} during eclipse ingress and eclipse. In contrast, Ne lines have not been reported before.

Soft residuals in emission remain visible in our observation (see hereafter  Sect.~\ref{section:strict_simult_xmm_nustar} and Fig.~\ref{fig:spec_strict_simult}). \citet{Rodes-Roca_2011} and \citet{Torrejon_2015} also reported the presence of a soft excess below 0.6 keV in the spectrum of \fouru with \xmmnewton and \chandra, respectively. Although adding a bremsstrahlung component improved their spectral fit, \citet{Rodes-Roca_2011} discarded this model due to the unrealistically high temperature it implied. They did not rule out a blackbody origin for the excess, which was the solution chosen by \citet{Torrejon_2015} with \chandra and \citet{Robba_2001} with \bepposax, although this would require an implausibly large accretion column or polar cap. They also tested a blend of Gaussian components, but this was less successful than modelling the continuum with three power laws, despite the difficulty in interpreting the physical origin of multiple power laws. 

The spectra of many X-ray pulsars, like, e.g., Her X-1, 4U~1626--67, Cen X-3, and Vela X-1, show a low-energy component, but its physical origin has remained a mystery \citep{Hickox_2004}. Given the lack of a widely accepted physical explanation for the soft excess, we choose to leave it unmodelled. Moreover, as seen in Fig.~\ref{fig:spec_strict_simult}, the soft excess in our data appears less pronounced than in \citet{Rodes-Roca_2011} and \citet{Torrejon_2015}. This may be due to the earlier orbital phase of our observation ($\phi_{\mathrm{orb}} \approx 0.55$), compared to their observations during eclipse ingress and eclipse, where the direct emission from the neutron star is more obscured and softer X-ray components become more dominant.

\subsection{Strictly simultaneous \xmmnewton and \nustar fitting}
\label{section:strict_simult_xmm_nustar}

Our final and best-fit model can be written as:
\begin{align*}
\label{eq:final_model}
\begin{split}
\mathcal{M}(E) =& \ \texttt{Tbabs}_{\rm{ISM}} \times \texttt{Tbpcf} \ \times \\ &(\texttt{Powerlaw} \times \texttt{Highecut} \times \prod^3_{i=1}\texttt{Gabs}_i \ + \sum^{3}_{j=1}\texttt{Gauss}_j)
\end{split}
\end{align*}

\begin{figure}
    \centering
    \centerline{\includegraphics[trim=0cm 0cm 0cm 0cm, clip=true, width=1.0\linewidth]{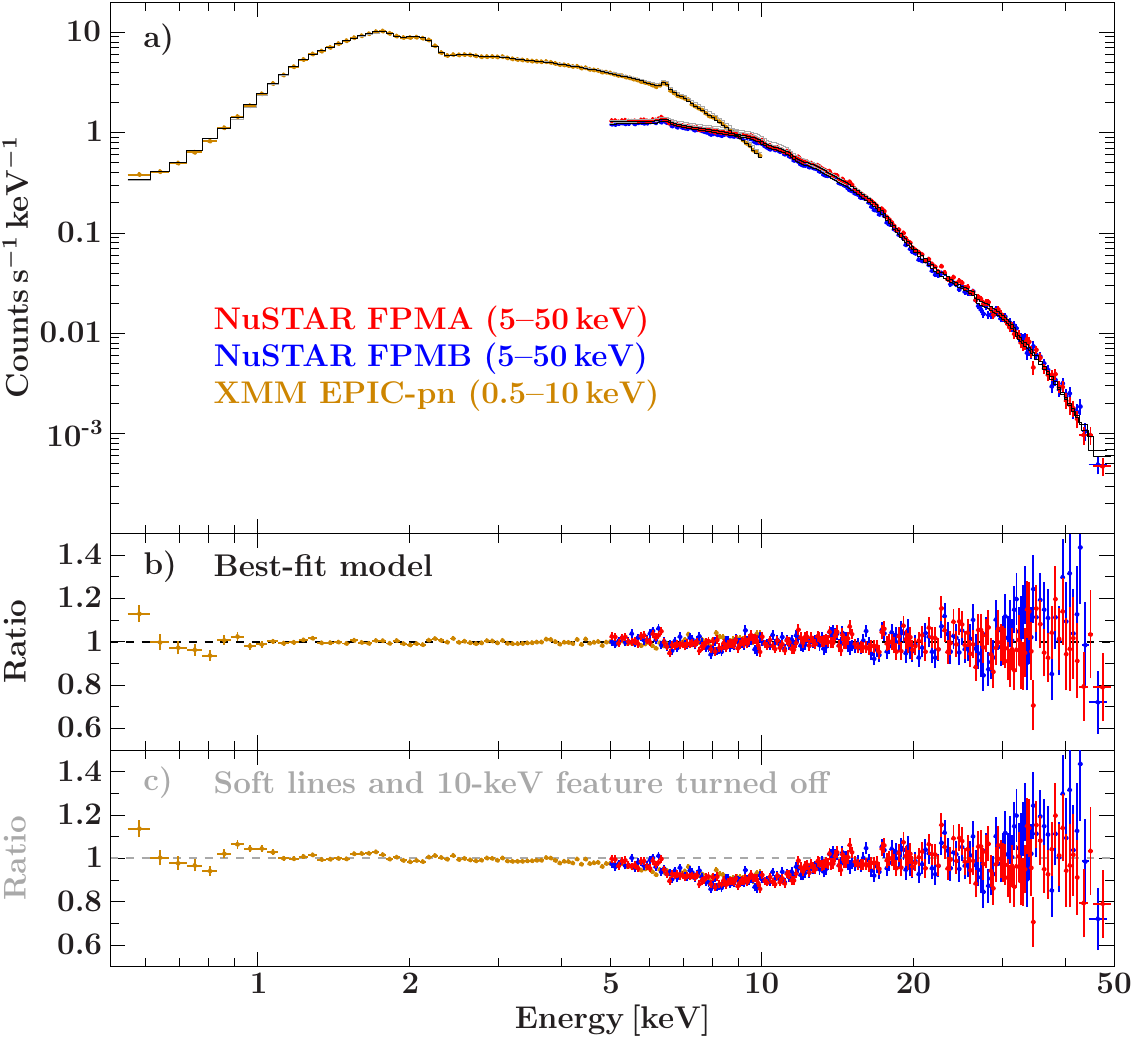}}
    \caption{\textit{Panel~a)}: Strictly simultaneous \xmmnewton/EPIC-pn (gold), \nustar/FPMA (red) and FPMB (blue) folded spectra. The black solid line corresponds to the best-fit model, and the grey solid line to the best-fit model with the soft lines and 10-keV feature turned off. \textit{Panel~b)}: Residuals for the best-fit model (black). \textit{Panel~c)}: Residuals for the best-fit model with the soft lines and 10-keV feature turned off (grey). The residuals in the lower panels are computed as data/model.}
    \label{fig:spec_strict_simult}
\end{figure}

We fit this model to the \xmmnewton and \nustar data when they are strictly simultaneous. Combining both datasets reduces the possible degeneracies between model parameters (e.g., the power law index with the absorption strength as seen in Fig.~\ref{fig:contour_plot} in the Appendix) and maximises the energy range covered. We present the best-fit parameters in Table~\ref{tab:best_fit_strict_simult}, the fitted folded spectra in \textit{panel a} of Fig.~\ref{fig:spec_strict_simult} and corresponding best-fit residuals in \textit{panel b}. We show the contribution from the soft lines in the Ne and Mg regions and from the broad 10-keV feature in \textit{panel c} of the same figure. Our continuum parameters are consistent with those found by \citet{Hemphill_2014}, who also used a \texttt{highecut} continuum with a partial coverer. The large covering fraction $\rm{CF} \approx 0.9$ is consistent with a neutron star deeply embedded in the wind of its stellar companion. However, the local low equivalent hydrogen absorption column density $N_{\rm{H,\,pcf}}$ value indicates that the local wind is relatively weak or not particularly dense. We measure an unabsorbed flux of $1.58^{+0.07}_{-0.08} \ \mathrm{keV\,s^{-1}\,cm^{-2}}$ or $2.54^{+0.10}_{-0.13}\times 10^{-9} \ \mathrm{erg\,s^{-1}\,cm^{-2}}$ in the 0.5--50 keV band. This corresponds to a mean intrinsic 0.5--50 keV luminosity of $9.56^{+3.42}_{-2.91} \times 10^{36} \, \mathrm{erg\,s^{-1}}$, assuming the distance (and errors) from \citet{Bailer_Jones21} of $\sim$5.6$^{+0.5}_{-0.4}$\,kpc. Our value lies at the higher end of the previously reported mean luminosities, which average around $10^{36} \, \mathrm{erg\,s^{-1}}$ \citep[e.g.,][]{Rodes-Roca_2011, Hemphill_2014, Hemphill_2019}, likely due to the prominent flare observed during this campaign. We also computed the 0.5--50 keV luminosity of the B-shaded region corresponding to the flare (see Fig.~\ref{fig:hr}). The derived value of $1.53^{+0.59}_{-0.47} \times 10^{37} \,  \mathrm{erg\,s^{-1}}$ is comparable with the flare observed by \suzaku in \citet{Hemphill_2014} measured at $1 \times 10^{37} \, \mathrm{erg\,s^{-1}}$ in the 5--100 keV band.

Regarding calibration constants, we note a flux offset of $\sim$20\% between EPIC-pn and both FPMs, consistent with what \citet{Diez_2023} measured when analysing simultaneous EPIC-pn and \nustar data of the HMXB Vela X-1. We additionally note a negative $\sim$50-eV energy offset of FPMB relative to FPMA and a positive $\sim$25-eV offset of EPIC-pn relative to FPMA, implying an absolute $\sim$75-eV offset between EPIC-pn and FPMB. \citet{Ballhausen_2020}, \citet{Ballhausen_2024}, and \citet{Zalot_2024} reported an offset of $\sim$40 eV, which corresponds to $\sim$1 \nustar energy bin, between both FPMs when analysing the HMXBs IGR J16318--4848, EXO 2030+375 and GX 301--2, respectively. This is also in agreement with estimated uncertainties derived in \citet{Grefenstette_2022} between FPMA and FPMB, and the energy shift of 87 eV reported by \citet{Diez_2023} between EPIC-pn and \nustar. We therefore conclude that our flux and offset constants (including uncertainties on the values) are within the expected ranges reported in the literature for other HMXBs.

\begin{table}[!htpb]
\caption{Best-fit parameters for the time-averaged strictly simultaneous \xmmnewton and \nustar spectra in Fig.~\ref{fig:spec_strict_simult}.}
\label{tab:best_fit_strict_simult}
\begin{center}
\begin{small}
\begin{tabular}{lll}    
\hline
\hline
\multicolumn{1}{l}{Parameter} & 
\multicolumn{1}{l}{Prior distribution} &
\multicolumn{1}{l}{Best-fit value}\\
\hline

$N_{\rm{H,ISM}} \ (10^{22} \, \rm{cm^{-2}})$  & 
fixed to 0.733 & - \\
$N_{\rm{H,\,pcf}} \ (10^{22} \,  \rm{cm^{-2}})$ &
$\mathcal{U}(0, \, 20)$ &  
$1.096^{+0.049}_{-0.048}$  \\
$\rm{CF}$  & $\mathcal{U}(0.7, \, 1)$ & $0.895^{+0.016}_{-0.017}$ \\
$\Gamma$  & $\mathcal{U}(0.5, \, 2)$ &   $1.045^{+0.024}_{-0.023}$ \\
$K \ (\rm{ph \ keV^{-1} \ cm^{-2} \ s^{-1}})$ & $\log \mathcal{U}(10^{-4}, \, 1)$ &  $0.0365 \pm 0.0010$  \\
$\mathcal{F}_{0.5-50\,\mathrm{keV}}^{(a)} \, \mathrm{(keV\,s^{-1}\,cm^{-2})}$ & - & $1.58^{+0.07}_{-0.08}$ \\
$E_{\rm{cut}} \ (\rm{keV})$  & $\mathcal{U}(8, \, 20)$ &  $16.99^{+0.40}_{-0.44}$  \\
$E_{\rm{fold}} \ (\rm{keV})$  & $\mathcal{U}(7, \, 25)$ &  $8.8^{+0.3}_{-0.4}$  \\
$E_{10\,\rm{keV}} \ (\rm{keV})$  &   $\mathcal{U}(8, \, 10)$ & $9.24^{+0.19}_{-0.20}$ \\
$d_{10\,\rm{keV}} \ (\rm{keV})$  &   $\mathcal{U}(0, \, 1)$ & $0.83^{+0.17}_{-0.19}$  \\
$\sigma_{10\,\rm{keV}} \ (\rm{keV})$  & $\mathcal{U}(0, \, 5)$ &  $2.60^{+0.23}_{-0.25}$  \\
$E_{\rm{CRSF,F}} \ (\rm{keV})$  &   $\mathcal{U}(10, \, 30)$ & $21.00 \pm 0.15$  \\
$\sigma_{\rm{CRSF,F}} \ (\rm{keV})$  &  $\mathcal{U}(0, \, 10)$ & $4.29 \pm 0.20$ \\
$d_{\rm{CRSF,F}} \ (\rm{keV})$  &   $\mathcal{U}(0, \, 30)$ & $8.6 \pm 0.7$  \\
$E_{\rm{CRSF,H}} \ (\rm{keV})$  &  $\mathcal{U}(40, \, 60)$ &  $ \geq  45.45$  \\
$\sigma_{\rm{CRSF,H}} \ (\rm{keV})$  &  fixed to $2\times \sigma_{\mathrm{CRSF,F}}$  & -\\
$d_{\rm{CRSF,H}} \ (\rm{keV})$  &  $\mathcal{U}(0, \, 5)$ &  $2.7^{+2.3}_{-2.4}$  \\
$E_{\rm{Ne}} \ (\rm{keV})$  &  $\mathcal{U}(0.85, \, 1.2)$ & $0.982^{+0.041}_{-0.046}$ \\
$A_{\rm{Ne}} \ (\rm{ph \, s^{-1} \, cm^{-2}})$  &  $\log \mathcal{U}(10^{-8}, \, 0.005)$ &  $0.51^{+0.41}_{-0.35} \times 10^{-3}$\\
$\sigma_{\rm{Ne}} \ (\rm{keV})$  &  $\mathcal{U}(10^{-6}, \, 0.1)$ & $0.056^{+0.044}_{-0.042}$  \\
$E_{\rm{Mg}} \ (\rm{keV})$  &  $\mathcal{U}(1.6, \, 2)$ &  $1.64^{+0.04}_{-0.03}$  \\
$A_{\rm{Mg}} \ (\rm{ph \, s^{-1} \, cm^{-2}})$  &   $\log \mathcal{U}(5\times10^{-6}, \, 0.005)$ & $0.17^{+0.08}_{-0.06} \times 10^{-3}$  \\
$\sigma_{\rm{Mg}} \ (\rm{keV})$  & $\mathcal{U}(0.001, \, 1)$ & $0.062^{+0.041}_{-0.046}$  \\
$E_{\rm{Fe\,K\alpha}} \ (\rm{keV})$  & $\mathcal{U}(6.3, \, 6.5)$ & $6.39 \pm 0.01$  \\
$A_{\rm{Fe\,K\alpha}} \ (\rm{ph \, s^{-1} \, cm^{-2}})$  & $\log \mathcal{U}(5\times10^{-6}, \, 0.005)$ & $0.18^{+0.02}_{-0.03}\times 10^{-3}$  \\
$\sigma_{\rm{Fe\,K\alpha}} \ (\rm{keV})$  & $\mathcal{U}(10^{-8}, \, 0.01)$ & $2.8^{+1.1}_{-1.5}\times10^{-3}$  \\
$\mathcal{C}_{\rm{FPMA}}$  &  fixed to 1  & -\\
$\mathcal{C}_{\rm{FPMB}}$  & $\mathcal{U}(0.5, \, 1.5)$ & $1.012^{+0.009}_{-0.010}$  \\
$\mathcal{C}_{\rm{EPIC-pn}}$ & $\mathcal{U}(0.5, \, 1.5)$ &  $0.792 \pm 0.005$ \\
$\mathcal{S}_{\rm{FPMA}} \ (\rm{eV})$  & fixed to 0  & -\\
$\mathcal{S}_{\rm{FPMB}} \ (\rm{eV})$  & $\mathcal{N}(0, \, 50)$ & $-50^{+42}_{-51}$  \\
$\mathcal{S}_{\rm{EPIC-pn}} \ (\rm{eV})$  &  $\mathcal{N}(0, \, 50)$ & $+25^{+13}_{-15}$  \\
C-stat/dof$^{(b)}$  & - & 654.32/379 \\
Reduced C-stat$^{(b)}$  & - & 1.73 \\
\hline
\multicolumn{3}{p{1.0\linewidth}}{$^{(a)}$ Unabsorbed flux.} \\ 
\multicolumn{3}{p{1.0\linewidth}}{$^{(b)}$ The parameters derived from our variational inference analysis with \texttt{jaxspec} were loaded in ISIS to compute the equivalent C-stat to allow for frequentist goodness of fit assessment.} \\ 
\end{tabular}
\end{small}
\end{center}
\end{table}

\section{Time-resolved spectroscopy}
\label{section:spectral_analysis}

\subsection{Per exposure cycle: \xmmnewton + \nustar}
\label{section:gti_spectro}

\begin{figure}
    \centering
    \centerline{\includegraphics[trim=0cm 0cm 0cm 0cm, clip=true, width=1.0\linewidth]{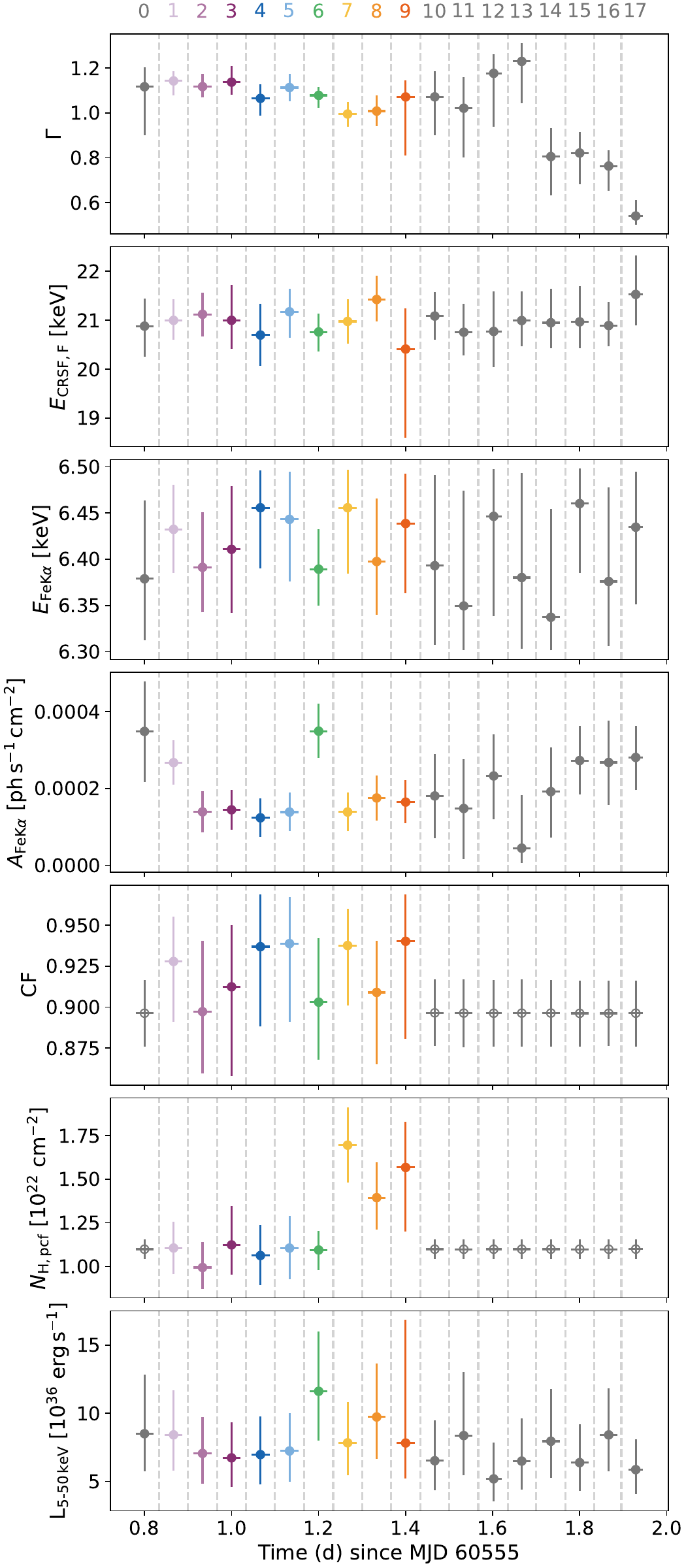}}
    \caption{Best-fit spectral parameters of interest per GTI as a function of time with \xmmnewton and \nustar. From top to bottom: photon index $\Gamma$, energy of the fundamental CRSF ($E_{\mathrm{CRSF,F}}$), energy of the Fe K$\alpha$ line ($E_{\mathrm{Fe\,K\alpha}}$), photon flux of the Fe K$\alpha$ line ($A_{\mathrm{Fe\,K\alpha}}$), covering fraction (CF), local absorption column density ($N_{\mathrm{H,\,pcf}}$), and unabsorbed luminosity in the 5--50\,keV band ($L_{\mathrm{5-50\,keV}}$). Coloured markers (from GTI 1 to GTI 9) show results obtained from overlapping \xmmnewton and \nustar data, and dark grey markers (GTI 0 and GTI $\geq 10$) from \nustar coverage only. Empty markers could not be constrained from \nustar alone and were fixed to their values from Table~\ref{tab:best_fit_strict_simult} derived from the time-averaged strictly simultaneous part of the observation.}
    \label{fig:params_vs_time_gti}
\end{figure}

To investigate shorter-timescale variability, we now perform time-resolved spectroscopy. For consistency of the results, we use the same spectral model defined in Sect.~\ref{section:strict_simult_xmm_nustar}. Because the background stays stable during the whole observation, we use the \nustar and \xmmnewton EPIC-pn time-averaged background spectra for all GTIs, as they provide a higher S/N. The LEO of \nustar provides natural GTIs for high-time-resolved spectroscopy. In Fig.~\ref{fig:hr}, we show each \nustar exposure cycle covered during this observation, indexed by a number and delimited by vertical dashed lines, from which we extract both \nustar and \xmmnewton data (when available). Due to the different start and stop times of the observations, only GTI 1 to GTI 9 benefit from a simultaneous \nustar and \xmmnewton coverage. For GTIs 0 and $\geq 10$, we have to fix $\mathrm{CF}$, $N_{\mathrm{H,pcf}}$, and the spectral parameters of the Ne and Mg features to the averaged values reported in Table~\ref{tab:best_fit_strict_simult}, as these parameters cannot be constrained using \nustar data alone. We note, however, that the uncertainties associated with these fixed parameters are still propagated using the errors derived from the time-averaged analysis. We present the results of the analysis in Fig.~\ref{fig:params_vs_time_gti}, focusing on parameters of interest. 

The photon index $\Gamma$ varies moderately in the 1.0--1.2 range from the start of the observation through GTI 11, albeit showing a local minimum during GTI 7 following the flare (see Fig.~\ref{fig:hr}). This trend is followed by a drastic decrease from GTI 14 through the end of the observation, reaching below 0.6. The energy of the fundamental cyclotron line is found between 20\,keV and 22\,keV, although we note local maxima at GTI 8 following the flare and during the last GTI of the observation. Dispersion in the Fe K$\alpha$ line energy are expected due to the velocity perturbations in the wind \citep[$\sim$200\,$\mathrm{km\,s^{-1}}$,][]{Diez_2025} and/or if the material is ionised \citep[up to $\sim$40 eV for low Fe K ionisation states,][]{Kallman_2004}. However, in Fig.~\ref{fig:params_vs_time_gti}, we observe a weakly constrained Fe K$\alpha$ line energy in each GTI around $\sim$6.4\,keV for which we are unable to identify any significant overall trend with time. This is possibly due to the limited energy resolution of \xmmnewton and \nustar combined with the large uncertainties in the relative energy shifts between the instruments, and the fact that the end of the observation is only covered by \nustar. 

Nonetheless, we note a high Fe K$\alpha$ line flux during GTI 6 that overlaps with the flare and a line flux consistent with 0 at GTI 13, coinciding with a local dip in the \nustar light curve. The $\mathrm{CF}$ remains stable at high values during the whole \xmmnewton observation, consistent with an absorbing material covering most of the line of sight. Although $N_{\mathrm{H,pcf}}$ increases by more than 50\% at GTI 7, immediately following the flare, the values remain low, further supporting the presence of a weak wind. It also does not appear to correlate with changes in spectral shape, contrary to what was observed in Vela X-1 \citep{Diez_2023}. 

We note that the covering fractions $\rm{CF}$ derived from the time-resolved spectroscopy are on average higher than the value obtained from the time-averaged strictly simultaneous \xmmnewton and \nustar spectra (Table~\ref{tab:best_fit_strict_simult}). This discrepancy is most likely related to the imperfect modelling of the low-energy upturn observed in \nustar (see Sect.~\ref{section:nustar}), which can affect continuum parameters describing the low-energy part of the spectrum. As a result, the $\rm{CF}$ inferred from the joint \xmmnewton--\nustar fit may be biased towards lower values compared to that derived from \xmmnewton alone. A similar issue was encountered by \citet{Diez_2023}, who adopted two distinct $\rm{CFs}$, one for each instrument, to account for calibration and modelling differences, and obtained $\rm{CF_{NuSTAR}} < \rm{CF_{XMM}}$. In our analysis, we try to minimise this effect by restricting the \nustar low energy threshold to 5 keV rather than using the standard 3 keV limit, and by adding two linear cross-calibration constants (see Sect.~\ref{section:spectral_modelling}).

\subsection{Pulse-by-pulse: \xmmnewton}
\label{section:pulse_spectro}

\begin{figure}[htpb!]
    \centering
    \centerline{\includegraphics[trim=0cm 0cm 0cm 0cm, clip=true, width=1.0\linewidth]{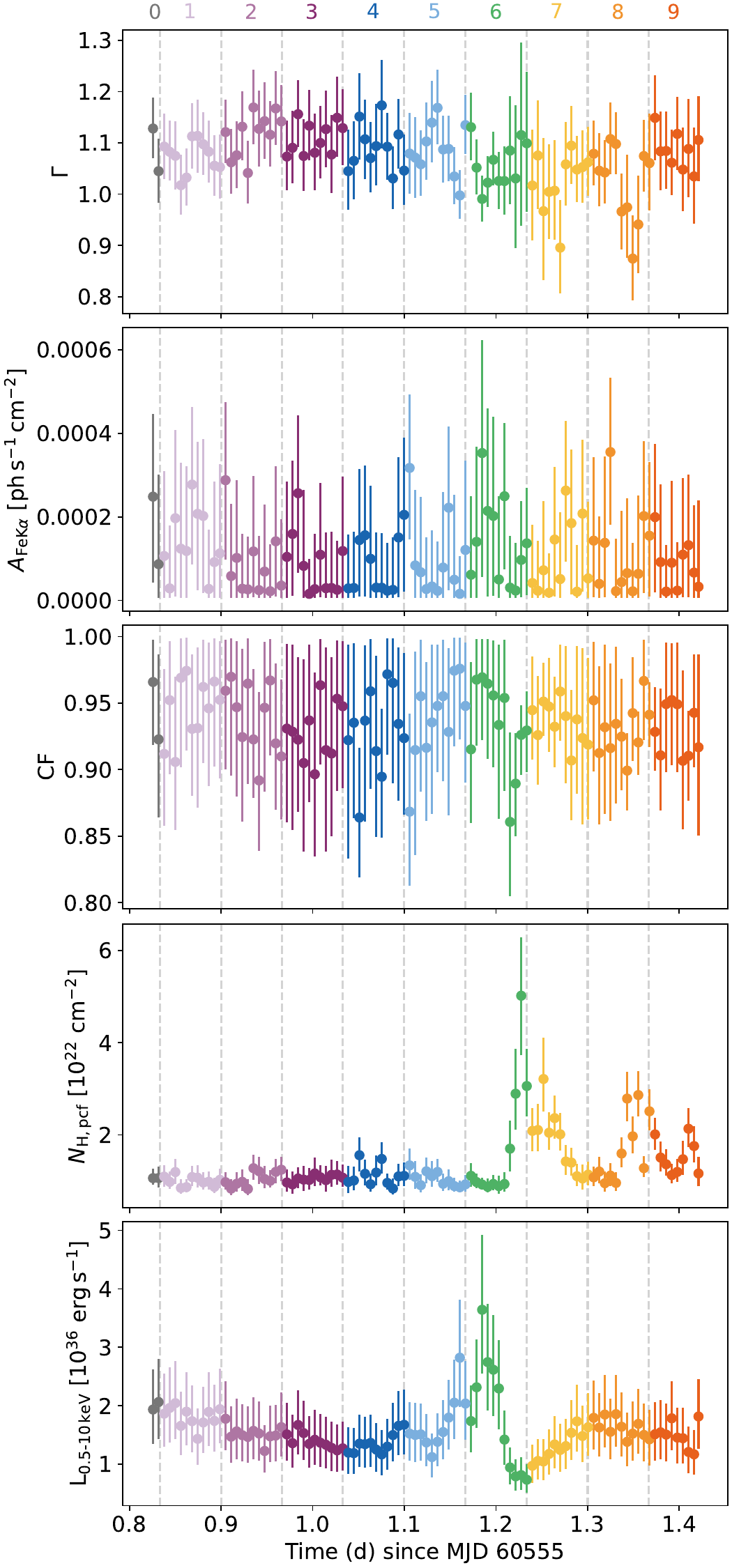}}
    \caption{Best-fit spectral parameters for the pulse-by-pulse time-resolved spectroscopy as a function of time using \xmmnewton only. From top to bottom: photon index $\Gamma$, photon flux of the Fe K$\alpha$ line ($A_{\mathrm{Fe\,K\alpha}}$), 
    covering fraction (CF), local absorption column density ($N_{\mathrm{H,\,pcf}}$), and unabsorbed luminosity in the \xmmnewton EPIC-pn energy band ($L_{\mathrm{0.5-10\,keV}}$). Coloured markers indicate the different GTIs (or \nustar exposure cycles) indexed at the top of the figure, following the notation adopted throughout the paper. The model parameters that are not shown here are frozen to their GTI best-fit value (see Sect.~\ref{section:pulse_spectro}).}
    \label{fig:params_vs_time_pulse}
\end{figure}

In this section, we conduct higher-time-resolved spectroscopy for a deeper investigation of the trends described in Sect.~\ref{section:gti_spectro}. Thanks to the high-time resolution permitted by \xmmnewton/EPIC-pn in small window mode, we extract spectra down to $\sim$526\,sec ($P_{\rm{spin}}$). However, due to a limited S/N, we do not extract \nustar at such time resolution. For the following, we only consider the \xmmnewton/EPIC-pn dataset. For this pulse-by-pulse spectroscopy, the prior distribution of each extracted spectrum is defined from the posterior distribution of the corresponding GTI spectrum derived in Sect.~\ref{section:gti_spectro}. Given a reduced S/N compared to the previous section and to avoid model degeneracies, we freeze all parameters to their best-fit value except parameters of interest, for which we can constrain their value. We thus thaw the photon index $\Gamma$, the photon flux of the Fe K$\alpha$ line $A_{\mathrm{Fe\,K\alpha}}$, the covering fraction $\rm{CF}$, the local absorption column density $N_{\mathrm{H,\,pcf}}$, and the unabsorbed luminosity $L_{\mathrm{0.5-10\,keV}}$ in the \xmmnewton EPIC-pn energy band and show the best-fit values in Fig.~\ref{fig:params_vs_time_pulse}.
The major difference with Fig.~\ref{fig:params_vs_time_gti} lies in the higher time resolution, enabling us to explore the variability of the parameters in far more detail. For instance, we note that the previously detected increase in $N_{\rm{H,\,pcf}}$ is due to three independent peaks reaching $\sim$5, $\sim$3, and $\sim$2 times the baseline level, respectively. The first and major $N_{\rm{H,\,pcf}}$ peak also coincides well with a narrow decrease in $\rm{CF}$ at the end of GTI~6 during the post-flare dip in luminosity and is consistent with the HR observed in \textit{panel c} of Fig.~\ref{fig:hr}. The two following $N_{\rm{H,\,pcf}}$ peaks happen simultaneously with two dips in $\Gamma$ visible during GTI~7 and 8. We discuss the physical origin of these trends in the following section.

\section{Discussion}
\label{section:discussion}

\subsection{Orbital dependence of the Fe K$\alpha$ emission}

We measure an average Fe K$\alpha$ line flux of $0.18^{+0.02}_{-0.03}\times 10^{-3}\,\rm{ph \, s^{-1} \, cm^{-2}}$ at $\phi_{\rm{orb}} \approx 0.5$ (inferior conjunction). This value is slightly higher than that reported by \citet{Torrejon_2015} at $\phi_{\rm{orb}} \approx 0.25$ (quadrature), although the two measurements remain consistent within their respective uncertainties and were obtained using different continuum models. \citet{Torrejon_2015} proposed that the majority of the Fe K$\alpha$ emission originates from regions close to the neutron star ($< 1\,R_{\star}$). Under this scenario, the line flux is expected to peak near inferior conjunction, when the observer has the most direct view of the illuminated wind. While our measurement is qualitatively consistent with this expectation, the increase compared to quadrature is relatively modest. Furthermore, the Fe K$\alpha$ line flux in \fouru is weaker than that reported for other wind-fed HMXBs such as GX~301$-$2 \citep[][at pre-periastron]{Fuerst_2011}, Vela~X-1 \citep[][also at inferior conjunction]{Diez_2025} or OAO 1657$-$415 \citep{Pradhan_2023}. This may be due to the comparatively low wind density in \fouru measured during our observation, which limits the fluorescence of Fe. Another possibility could be a high degree of ionisation of the circumstellar material, supported by the high observed luminosity of $\sim$10$^{37}\,\rm{erg\,s^{-1}}$ and the local increase in Fe K$\alpha$ line flux during the flare. This would reduce the efficiency of fluorescent line production. Such a scenario could be investigated with future monitoring campaigns covering different orbital phases within a single orbit. 

\subsection{Observational evidence of a trailing structure}

A gradual increase in the \nustar HR is observed in \textit{panel f} of Fig.~\ref{fig:hr}. A similar behaviour was reported for Vela X-1 during two distinct \nustar campaigns close to inferior conjunction and using comparable energy bands \citep{Diez_2022}. For \fouru this spectral hardening is accompanied by a decrease in the photon index $\Gamma$, particularly after the flare around GTI 7 and towards the end of the observation, as shown in the top panel of Fig.~\ref{fig:params_vs_time_gti}. In addition, this decrease in $\Gamma$ coincides with higher absorption, whereas the model degeneracy would predict a positive correlation between the two parameters (see Fig.~\ref{fig:contour_plot} in the Appendix). The observed hardening therefore likely has a physical origin. A possible explanation is the presence of a trailing overdense structure behind the neutron star, often interpreted as an accretion wake or trailing wake, along the line of sight, as already reported for this source in \citet{Mukherjee_2006} and \citet{Rodes-Roca_2015}. This is further evidenced in our hydrodynamic simulation (see Fig.~\ref{fig:3Dsimu} and corresponding Sect.~\ref{section:3Dhydro}) where we can see the formation of a large overdense structure in the extended part of the neutron star wake. It should be noted, though, that the decrease in $\Gamma$ observed may be enhanced due to a fixed $N_{\rm{H,\,pcf}}$. Allowing $N_{\rm{H,\,pcf}}$ to vary was not possible with the \nustar data alone and would have introduced large uncertainties owing to the strong positive model degeneracy between $N_{\rm{H,\,pcf}}$ and $\Gamma$ (see Fig.~\ref{fig:contour_plot}).

\subsection{Comparison to 3D Hydrodynamics}
\label{section:3Dhydro}

\begin{figure}
    \centering
    \centerline{\includegraphics[trim=0cm 0cm 0cm 0cm, clip=true, width=1.0\linewidth]{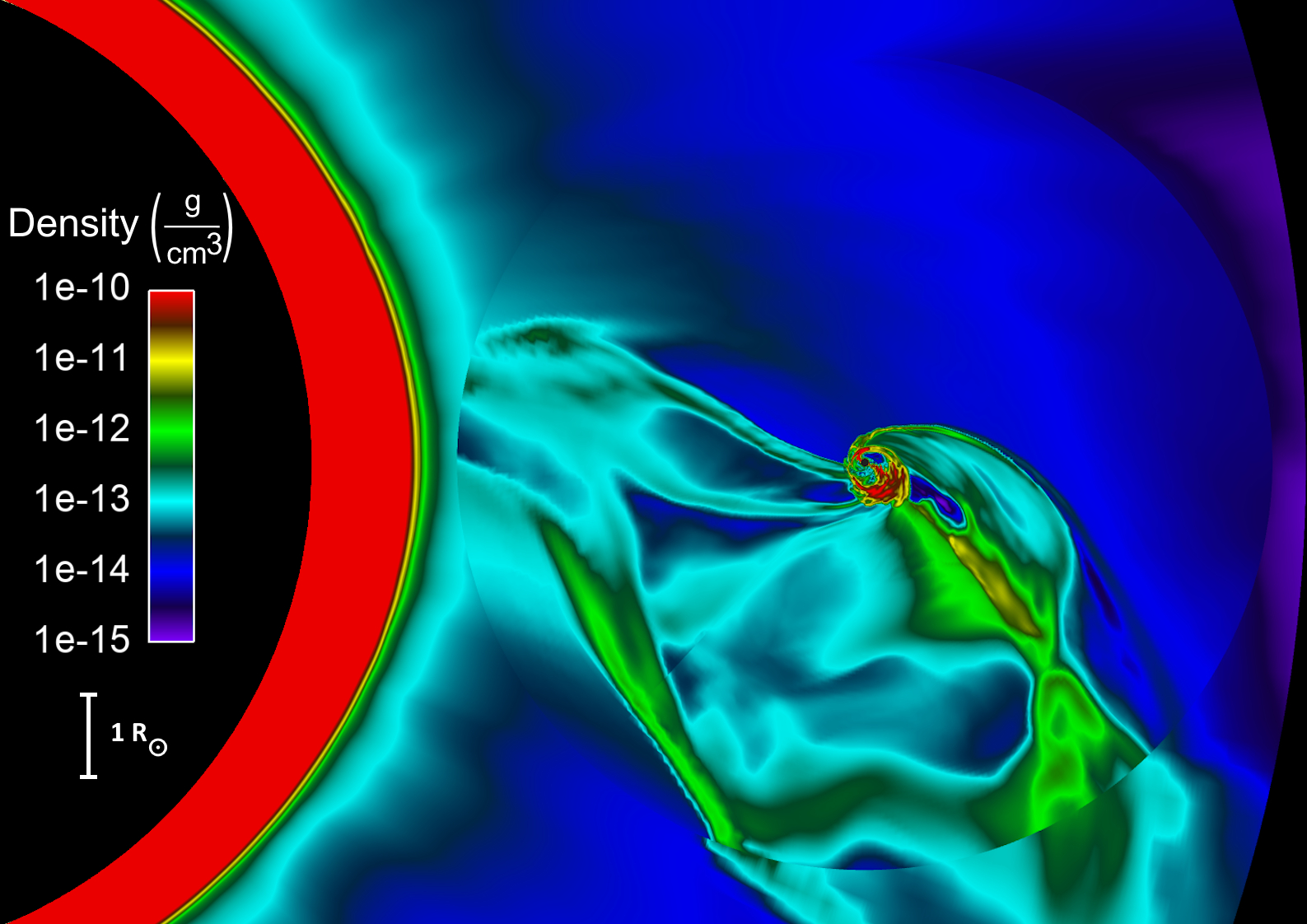}}
    \caption{Equatorial plane of a 3D hydrodynamic simulation of \fouru. Different volume densities are indicated by different colours as shown in the legend. We used a higher-resolution grid for regions near the neutron star, as highlighted by the circled region. The observer is facing the system on the right of the figure.}
    \label{fig:3Dsimu}
\end{figure}

We compare our results with a simulated light curve from a 3D hydrodynamic model of the same system (see Fig.~\ref{fig:3Dsimu}). The simulation is not intended to reproduce the observations quantitatively or to provide a best-fit model, but rather to illustrate the expected behaviour and enable a qualitative comparison with the observed light curve. We employ the hydrodynamics code VH-1 to model the entire binary system \citep{Blondin_2009}. We modify the method of \cite{Dickson_2024} by the addition of isotropic X-ray feedback. This approach dynamically varies the radiation pressure localised around the compact object with the rate of accretion (Guerriero \& Dickson, in preparation). Our simulation is initialised from the system parameters of \cite{Reynolds_1992} and \cite{Hemphill_2019} with an initially smooth steady state. 

\begin{figure}
    \centering
    \centerline{\includegraphics[trim=1.5cm 0.5cm 2.5cm 2.1cm, clip=true, width=1.0\linewidth]{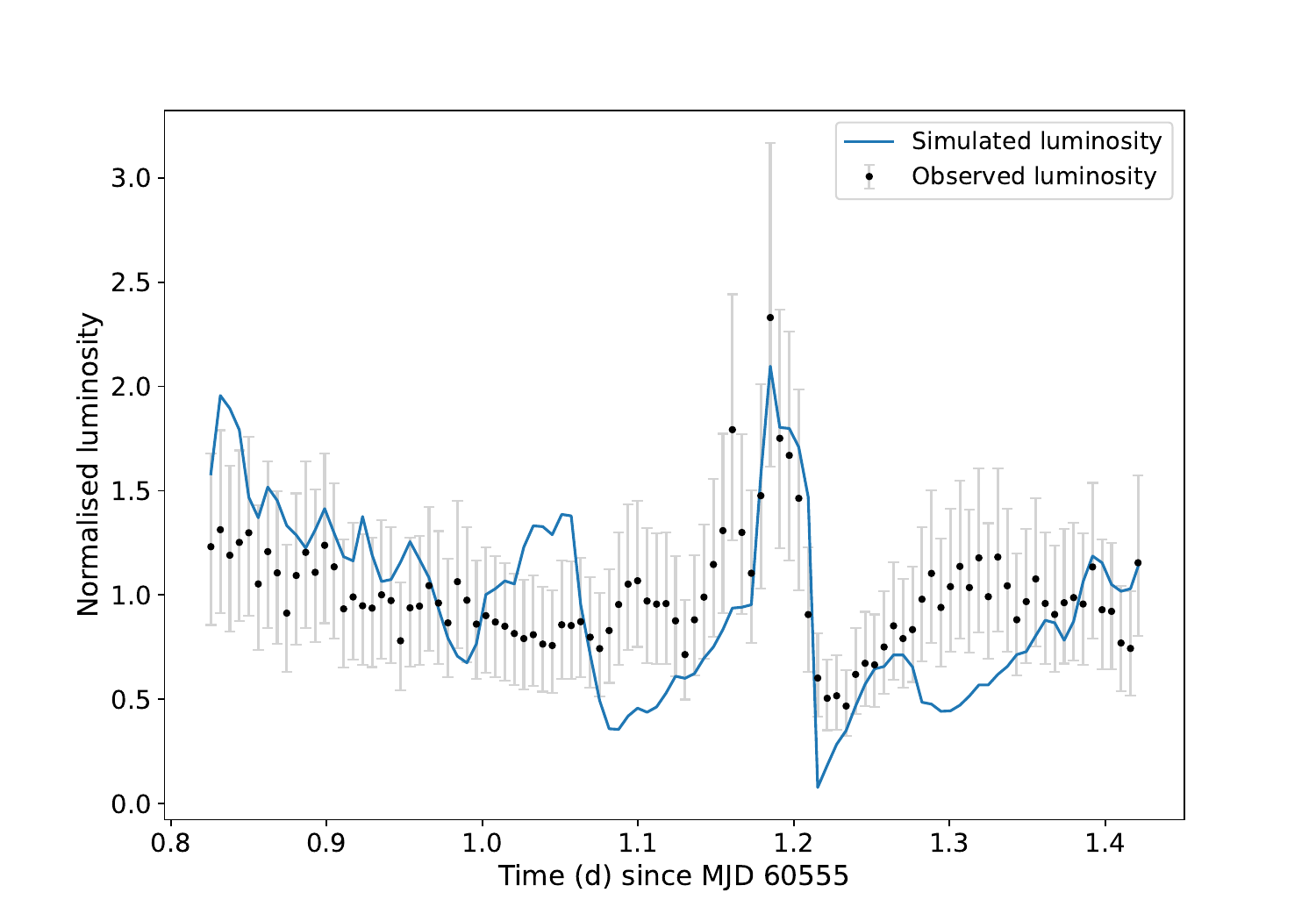}}
    \caption{Observed luminosity from Fig.~\ref{fig:params_vs_time_pulse} and simulated luminosity derived in Sect.~\ref{section:3Dhydro}, normalised to the average luminosity.}
    \label{fig:obssimu_lum}
\end{figure}

Our simulated light curve forms a luminosity peak comparable to the measured one (Fig.~\ref{fig:obssimu_lum}). The simulated peak results from an accreted overdensity within the wind outside the line of sight. Using the observed peak luminosity of $10^{37}\,\mathrm{erg\,s^{-1}}$ and a flare duration of approximately $t_{\mathrm{flare}}\approx12$ neutron star spin periods (Fig.~\ref{fig:params_vs_time_pulse} bottom panel), we estimate the mass of the accreted overdensity responsible for the flare as  $m_{\mathrm{acc}}=\frac{R_{\mathrm{NS}}}{GM_{\mathrm{NS}}}L_{\mathrm{flare}}t_{\mathrm{flare}} \approx 10^{20}\,\mathrm{g}$. In the hours following the luminosity peak, both the simulated and observed light curves display a dip below the continuum luminosity. This is consistent with an accretion stall caused by the temporarily heightened radiation pressure from the accretor. As the accretion rate rises, so does the corresponding radiation pressure, deflecting the donor wind and slowing its accretion. A stall in accretion would induce an accumulation of wind density in some sightlines in the neutron star vicinity, depending non-trivially on the path of deflected wind. This could account for the $N_{\mathrm{H,\,pcf}}$ peak we observe after the luminosity flare, though we do not detect an analogous feature in the simulated dataset. 

Furthermore, the observed peak in $N_{\rm{H}}$ coincides with a local decrease in the covering fraction $\rm{CF}$ (Fig.~\ref{fig:params_vs_time_pulse}) and a temporary hardening of the spectrum, as indicated by the HR (\textit{panel c} in Fig.~\ref{fig:hr}). This behaviour also favours a scenario in which the X-ray source is obscured by a geometrical structure, associated with a focused overdensity of material (often interpreted as a large clump or filamentary structures) crossing the line of sight.

\subsection{Constraints on clump properties from $N_{\rm{H}}$ variability}

To investigate the origin of the repeated peaks in $N_{\rm{H,\,pcf}}$ in Fig.~\ref{fig:params_vs_time_pulse}, we explore whether they can be produced by the same large absorber crossing the line of sight multiple times. We assume two clumps following circular Keplerian orbits at constant orbital velocity around the neutron star: one with an orbital period of 20 pulses ($\sim$0.12\,d) and the other one of 9 pulses ($\sim$0.055\,d) corresponding to the observed time intervals between the first and second $N_{\rm{H,\,pcf}}$ peak, and between the second and last $N_{\rm{H,\,pcf}}$ peak respectively. We also assume that the duration of each obscuration (12 pulses $\sim0.073$\,d, and 9 pulses $\sim 0.055$\,d) corresponds to the time during which the large clumps obscure the line of sight at their orbital velocity. Under these assumptions, the derived clump radius would exceed its orbital radius, which is clearly unphysical. This simple estimate rules out the interpretation that a single orbiting clump is responsible for the successive $N_{\rm{H,\,pcf}}$ peaks. Instead, they are more likely produced by distinct overdense structures embedded in the wake of the neutron star. This interpretation is further supported by our 3D hydrodynamic simulation (see Sect.~\ref{section:3Dhydro}) and by the shorter time interval between the second and third peak ($\sim$0.05\,d), which does not suggest any clear periodicity.

We perform a similar exercise for the pulse-to-pulse variability in $N_{\rm{H,\,pcf}}$, which is more clearly seen in the HR evolution (Fig.~\ref{fig:hr}). Although such short-timescale variations could be caused by individual clumps crossing the line of sight, they cannot be attributed to a single orbiting clump, as this would again imply unphysical parameters. Instead, these absorption events shorter than one neutron star pulse ($< P_{\rm{spin}}$) are more likely produced by distinct smaller clumps distributed along our line of sight. Following a similar method to that applied in \citet{Martinez_2014} for Vela X-1, we calculate the mass of a single clump causing such obscuration on the line of sight. Assuming a spherical clump located in the neutron star vicinity at a distance $d$, characteristic obscuration time of one pulse $\Delta t$ and clump velocity as the bulk of the wind $v$, we get:

\begin{equation*}
    l_{\mathrm{clump}} = 2\,R_{\mathrm{clump}} \approx v(d) \,\Delta t = v_{\infty}\left(1-\frac{R_{\star}}{d}\right)^{\beta}\,\Delta t \approx 10^{10}\,\mathrm{cm}
\end{equation*}
with the terminal wind velocity $v_{\infty} = 1400$--$2800\,\mathrm{km\,s^{-1}}$ \citep{Abbott_1982, Rodes-Roca_2015}, $d = 1.35$--$1.5\,R_{\star}$ \citep{Torrejon_2015} and $\Delta t = 526.03\,\mathrm{s} = P_{\rm{spin}}$. Assuming a constant density for the clump during the obscuration, we have $n_{\mathrm{H,\,clump}} = N_{\rm{H,\,peak}}/l_{\mathrm{clump}}$ with $N_{\rm{H,\,peak}} \approx 5\times 10^{22}\,\mathrm{cm^{-2}}$ for the primary peak. With a characteristic volume $V_{\mathrm{clump}}$ of $\frac{4}{3}\pi R_{\mathrm{clump}}^3$, we can derive the mass of the obscuring clump:

\begin{align*}
    m_{\mathrm{clump}} &= n_{\mathrm{H,\,clump}} \,V_{\mathrm{clump}} = \frac{\pi}{6} \, N_{\rm{H,\,peak}}\,l_{\mathrm{clump}}^2 \approx 10^{20}\,\mathrm{g}
\end{align*}

Our estimated clump mass of $\sim$10$^{20}\,\mathrm{g}$ is one to two orders of magnitude lower than the value reported by \citet{Hemphill_2014} for \fouru, but is consistent with the clump mass inferred for Vela X-1 by \citet{Martinez_2014}, who also observed a major flaring episode. Furthermore, our inferred clump density of $n_{\mathrm{H,\,clump}} \approx 10^{-12} \, \mathrm{g\,cm^{-3}}$ corresponds to a local overdensity of approximately three orders of magnitude relative to the ambient stellar wind density derived by \citet{Clark_1994}. Such density contrasts are also evident in our 3D hydrodynamic simulation (see Fig.~\ref{fig:3Dsimu}).

\subsection{Evolution of cyclotron line energy}

Observationally, the sub-critical (resp. super-critical) regime is often associated with a positive (resp. negative) correlation between cyclotron line energy and luminosity. Different models have been proposed to explain such correlations with luminosity \citep[see reviews of the field in][]{Mushtukov_Tsygankov_2022, Saathoff_PhD}, but the observational constraints remain too weak to select a preferred scenario. \citet{Hemphill_2016, Hemphill_2019} performed a time- and luminosity-resolved study of \fouru and did not find a clear CRSF energy--luminosity trend using \rxte, \integral, \suzaku and \nustar data. 

In our dataset, we cannot identify any clear trend between these two parameters, nor any obvious time-dependent evolution (see Fig.~\ref{fig:contour_plot} in the Appendix). Moreover, when converting to luminosity space, we introduce additional systematic uncertainties given the 10\% error in the source distance derived in \citet{Bailer_Jones21}. Consequently, we do not detect any significant correlation between $E_{\mathrm{CRSF,F}}$ and luminosity. That result, in alignment with previous studies \citep[e.g.,][]{Hemphill_2016, Hemphill_2019}, may suggest either that \fouru spends most of its observed time near a regime where the expected $E_{\mathrm{CRSF}}$--luminosity correlation is weak, or that the accretion geometry and/or long-term magnetic field configuration changes. \citet{Hemphill_2016} reported that $E_{\mathrm{CRSF,F}}$ followed a long-term increasing trend from the early 2000s to the 2010s. More recently, \citet{Tamang_2024} (see their Figure 9) suggested a possible decrease in the line energy since the 2010 \integral observation analysed by \citet{Hemphill_2016}. Our measured average value of $21.00\pm0.15$ keV seems to support this declining trend. However, as emphasised by \citet{Hemphill_2019}, the large uncertainties, the varying flux states, and different orbital phases covered strongly limit any conclusions regarding such correlations and their physical interpretation.

\section{Conclusion and outlook}
\label{section:conclusion}

In this work, we analyse the first out-of-eclipse \xmmnewton observation of the HMXB \fouru, complemented by a simultaneous \nustar observation. Our campaign covers the 0.5--50 keV energy range around inferior conjunction, providing a unique opportunity to investigate the structure of the accreted material in the vicinity of the neutron star. Using a partial-covering spectral model and the \texttt{jaxspec} analysis software within a Bayesian framework, we perform time-resolved spectroscopy down to the neutron star pulse period ($\sim$526 s) during a bright heartbeat-shaped flare reaching $\sim$10$^{37}\,\mathrm{erg\,s^{-1}}$.

We investigate the accretion regime of \fouru following the evolution of the cyclotron line features. As in previous studies, we do not find any significant correlation between the CRSF energy and the source luminosity. The limited S/N, particularly at high energies, together with parameter uncertainties, prevents us from drawing firmer conclusions on the accretion regime. Our analysis also reveals overdense structures on several spatial scales. The gradual hardening of the HR towards the end of the observation, mostly constrained by \nustar, probes the presence of a large-scale trailing overdense structure behind the neutron star. Following the flare, we detect three distinct peaks in the line-of-sight absorption column density, which we interpret as local overdense or filamentary structures forming in the accretion wake. Superimposed on these events, shorter-timescale absorption variability on the pulse-period timescale is consistent with independent/distinct clumps crossing the line of sight. We estimate a characteristic clump mass of $\sim$10$^{20}\,\mathrm{g}$, in agreement with previous estimates for wind-fed HMXBs. These observational results are supported by our 3D hydrodynamic simulations, which reproduce both the heartbeat-shaped flaring episode and the formation of overdense structures around inferior conjunction. Finally, we search for periodicity in the observed absorption peaks but found no significant evidence for a recurring orbiting absorber. Additional observations with \xrism at inferior conjunction will provide the spectral resolution required to investigate the ionisation structure of the accretion wake and further test the results obtained in this work.

\begin{acknowledgements}
CMD acknowledges support through the European Space Agency (ESA) Research Fellowship Programme in Space Science and the Centre National d'Etudes Spatiales (CNES) Postdoctoral Fellowship. SD and CMD acknowledge the support of CNRS/INSU and CNES. This project was provided with HPC and storage resources by GENCI at IDRIS, thanks to the grant 2025-AD010416032R1, on the supercomputer Jean Zay’s A100/H100 partitions. ALD acknowledges support through the ESA SCI-E Internship Programme.
SMN and JMT acknowledge support from grant PID2024-155779OB-C31 funded by MICIU/AEI/MICIU/AEI/10.13039/501100011033 and by “ERDF/EU”. 
 
\end{acknowledgements}

\bibliographystyle{macros/bibtex/aa}
\bibliography{references}

\begin{appendix}

\onecolumn

\section{Contour plot for the time-resolved spectroscopy based on the \nustar exposure cycle}
\label{appendix:contour_gti}

\begin{figure}[htbp!]
    \centering
    \includegraphics[trim=1cm 0.5cm 1.5cm 2.5cm, clip=true, width=1.0\linewidth]{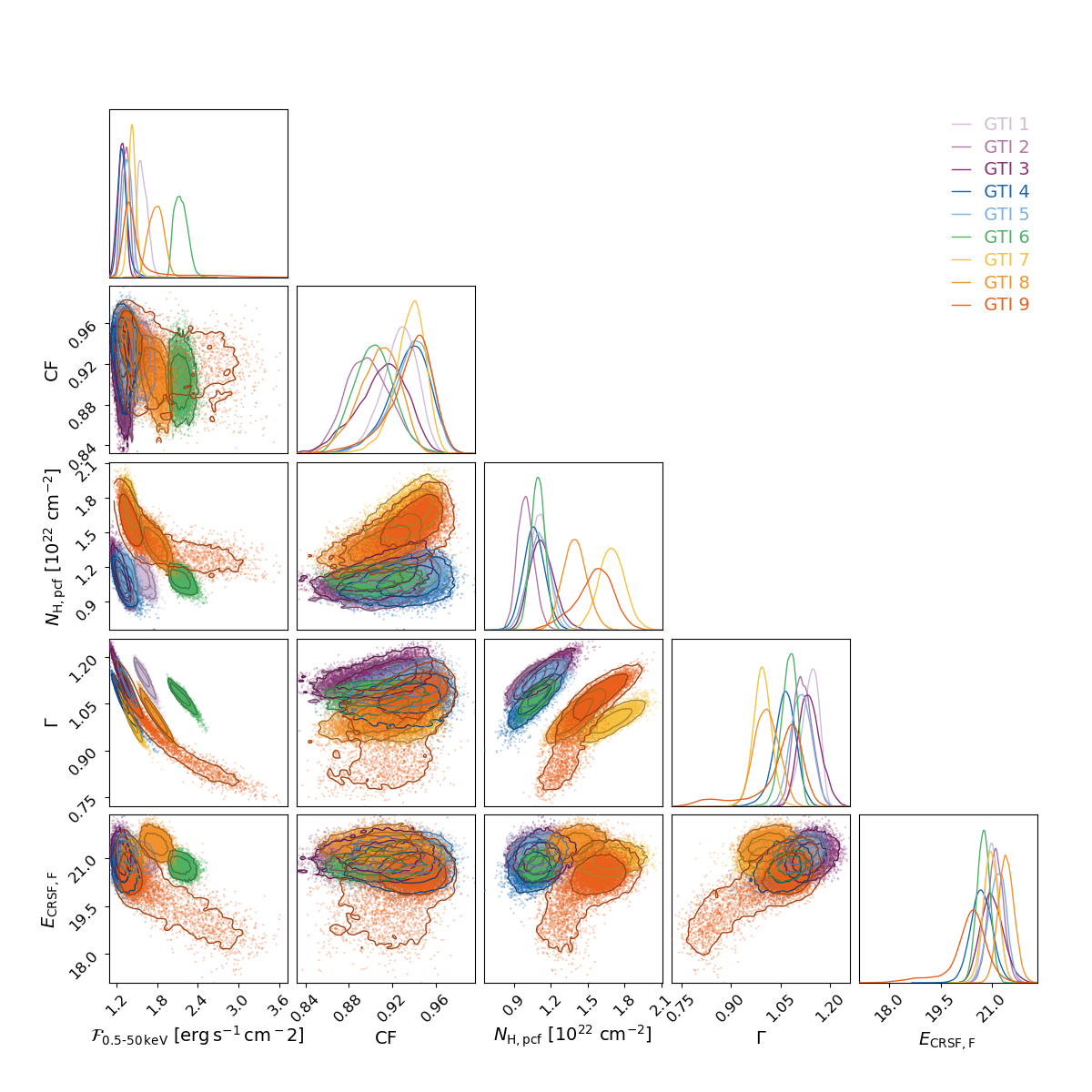}
    \caption{Contour plots of selected parameters derived from the MCMC sampling of our time-resolved spectroscopy. Different colours indicate the different GTIs we used based on the same colour code as in Fig.~\ref{fig:params_vs_time_gti}.}
    \label{fig:contour_plot}
\end{figure}

\twocolumn

\end{appendix}

\end{document}